\documentclass[twocolumn,superscriptaddress, prb, a4paper]{revtex4-1}
\usepackage{graphicx, amsmath, amssymb, subfigure, verbatim, appendix}
\usepackage{graphicx, subfigure,verbatim,amsmath, rotating}
\usepackage{hyperref}
\newcommand{\vect}[1]{\mathbf{#1}}
\newcommand{\eps}{\varepsilon}
\newcommand{\rl}{\rangle}
\newcommand{\lal}{\langle}
\begin{document}
\title{Optical recombination of biexcitons in semiconductors}
\author{M.~Bauer}
   \email{msb50@cam.ac.uk}
   \affiliation{Cavendish Laboratory, J~J~Thompson Avenue, Cambridge, CB3 0HE, United Kingdom}
\author{J.~Keeling}
   \affiliation{Scottish Universities Physics Alliance, School of Physics and Astronomy, University of St Andrews, St Andrews KY16 9SS, United Kingdom}
\author{M.~M.~Parish}
   \affiliation{London Centre for Nanotechnology, Gordon Street, London, WC1H 0AH, United Kingdom}
   \affiliation{Cavendish Laboratory, J~J~Thompson Avenue, Cambridge, CB3 0HE, United Kingdom}
\author{P.~L\'opez~R\'\i os}
   \affiliation{Cavendish Laboratory, J~J~Thompson Avenue, Cambridge, CB3 0HE, United Kingdom}
\author{P.~B.~Littlewood}
   \affiliation{Cavendish Laboratory, J~J~Thompson Avenue, Cambridge, CB3 0HE, United Kingdom}
   \affiliation{Physical Sciences and Engineering, Argonne National Laboratory, Argonne, Illinois 60439, U.~S.~A.}
\begin{abstract}
We calculate the  photoluminescence spectrum and lifetime of a biexciton in a semiconductor
 using Fermi's golden rule. Our biexciton wavefunction is obtained using 
a Quantum Monte Carlo calculation. For a recombination process where one 
of the excitons within the biexciton annihilates,
we find that the surviving exciton is most likely to populate the ground state.
We also investigate how the confinement of excitons in a
quantum dot would modify the lifetime in the limit of a large quantum
dot where confinement principally affects the centre of mass
wavefunction. The lifetimes we obtain are in reasonable agreement with
 experimental values.  Our calculation 
 can be used as a benchmark for comparison with approximate methods.
\end{abstract}
\maketitle

\section{Introduction}
Information about many-body systems of electrons and holes is primarily obtained
through their luminescence. While good calculations on the ground states of these systems
have been performed for a variety of electron and hole densities, good estimates of 
luminescence require the inclusion of final excited states, for which calculations
are difficult. Thus, it is important to quantify how excited final states affect 
luminescence. 
 In this paper, we solve essentially exactly the problem of decay
of the simplest correlated system, namely the biexciton.

Luminescence of these biexcitons, which are bound states of excitons (electron-hole pairs), 
is potentially relevant for studying luminescence from semiconductor lasers. 
While semiconductor lasers correspond to the dense limit of the electron-hole phase diagram,
 where the electrons and holes ionise into a plasma \cite{littlewoodbes02}, 
it is well known that the inclusion of correlation effects with the surrounding electrons 
and holes is necessary for a good estimate of semiconductor 
luminescence \cite{haugh67, haugkoch, schmittrinkeh86}. 
Moreover, a variational Monte Carlo study of the electron-hole phase diagram has 
revealed a surprisingly large excitonic insulator phase, which extends well 
into these high densities \cite{zhulittlewood96}.
Thus, luminescence from excitonic states also contributes to light
emission in the plasma. However, it is not a priori obvious whether 
emission from an independent exciton can adequately describe emission 
from a plasmonic state  where interaction effects are crucial. This
question of interaction can be addressed by considering the biexciton as
the most complicated correlated state that can still be treated exactly.
Since the detailed form of the bound state wavefunction can
significantly affect the overlap with final states it is important to go beyond
the use of simple analytical variational wavefunctions for the biexciton.  Instead
we use a  biexcitonic wavefunction obtained from a quantum Monte Carlo calculation. 
Thus, our results can be used as a benchmark for larger and more complicated systems 
that currently still require approximate methods.

The concrete aim of this work is twofold: we primarily want to calculate the
emission spectrum of the biexciton, using an essentially numerically
exact biexciton wavefunction. We then want to find out how well the
lifetime we calculate from this spectrum compares to experimental and
other theoretical lifetimes.

For the calculation of the lifetime we focus on radiative recombination involving
the emission of a photon. This process could potentially excite the surviving exciton into an
excited state. In order to obtain both the emission spectrum and the lifetime we
follow the procedure by Elliott \cite{elliott57} and use Fermi's golden rule, which requires
the calculation of matrix elements between these exciton excited state and the full
biexciton wavefunction. The wavefunctions and matrix elements are clearly different for
different mass ratios. Hence, we consider different mass ratios where the hole mass $m_h$ is equal to
or much greater than the electron mass $m_e$ for the spectra and several intermediate
mass ratios for the calculation of the lifetime.
 
It has been shown in previous calculations \cite{Ivanov} that the inclusion of the accurate
semiconductor eigenstates, i.e. polariton states, is important
for an accurate estimate of the biexciton lifetime. 
The importance of polariton effects for the biexciton decay
comes from the fact that since most binding energies of excitonic molecules
 $\eps_m$ are much smaller than the polariton splitting parameter $\Omega_c$, 
the dispersion of the final states after the decay process is strongly modified compared
to the dispersion of the free exciton and photon states. 
Hence we estimate lifetime including polariton effects; in doing this we use insights from our
calculation in the absence of polariton effects in order to simplify the
calculation in presence of polaritons. We present 
a general formula  
of the typical non-radiative biexciton lifetime in the semiconductor, which should 
in principle work for materials where the polariton effect is weak. 

To the extent that our Monte Carlo biexciton wavefunction is a good 
wavefunction, the remainder of our calculation is essentially exact for system where
the polariton effect is negligible. We expect that shape of our emission spectra, calculated
without polariton effects, will not be affected drastically even for materials
where the effect is important. 
Thus, our calculation can be used
as a comparator for approximate methods, which could, however, reach non-zero densities. 


While our spectra are for 3D bulk only, we extend our lifetime calculation to a 3D 
quantum dot, since more experimental data is available there.  We compare our lifetime
estimates for a range of materials to experimental and other theoretical estimates.
 A similar procedure for the calculation of lifetimes as presented in this paper
 may also be applied to quantum well systems.

In this paper we calculate the simplest optical recombination process.
The question of whether this is in fact responsible for the exciton
lifetime is however more complicated as many other mechanisms can play a
role. In semiconductor bulk systems these mechanisms for example 
 include recombination involving impurities, or traps \cite{shockleyr52},  
or Auger recombination \cite{haug78, pincherle55, beattiel59,lius06,kavoulakisbw96, kavoulakisb96,jolkjk02}. We discuss some of these later; however as our lifetimes compare reasonably with
experimental values it seems that radiative recombination is the main channel for
biexciton annihilation.
 
The paper is organised as follows:  Section II discusses photoluminescence from
biexcitons in a semiconductor, and section III discusses their lifetime. In 
section \ref{sec:s1m} we introduce the approach used to
determine the biexciton wavefunctions and how to calculate photoluminescence spectrum,
which we present and discuss in section \ref{sec:s1r}. Section \ref{sec:s2m} presents
the calculation of the biexcitonic lifetime with polariton effects and optionally confinement
 and section \ref{sec:s2r}
compares biexciton lifetimes for five semiconducting systems to experimental and some
other theoretical values, and discusses their trends with electron/hole mass ratio and confinement.
We conclude in section \ref{sec:concl}.

\section{Photoluminescence}
\subsection{Methods \label{sec:s1m}}
In this section we present the calculation of luminescence from the biexciton.
We first discuss the methodology of this calculation, starting from the light-
matter interaction Hamiltonian, and the form of the biexcitonic wavefunction, before
quoting the general formulae for the relative transition rates into different excited
excitonic states. We then present the results and discuss our obtained emission spectra.

The light-matter interaction in second quantisation is given by
\begin{align} \label{eq:h}
H' = \gamma &\times \int  dr \sum_{\sigma' = \pm}
     \sum_{\vect{k}_e,\vect{k}_h \in BZ} 
     \frac{e}{m_0} 
 \quad A_0 \nonumber \\& a_{\vect{k}_e+\vect{k}_h,\sigma'}^\dagger
 c_{-\vect{k}_h,\sigma}^\dagger c_{\vect{k}_e,-(\sigma -\sigma')}
p_{cv}
\end{align}
where $p_{cv}$ is the momentum matrix element for a transition of an electron
with charge $e$ and mass $m_0$ between the valence and the conduction band and 
$A_0=\sqrt{\hbar/(2 \epsilon \omega_k V)}$ is the standard vector potential field strength.
This expression contains $\omega_k$ as the frequency of the emitted photon, the dielectric
constant of the material $\epsilon$, and a unit of box quantisation volume $V$  which cancels in 
the latter part of the calculation.
The operators  $c_{\vect{k}_{h/e}, \sigma }^\dagger$ create holes and electrons
with momentum $k_h$ and $k_e$ and with spin quantum number
$\sigma =  \pm \frac{3}{2}, \frac{1}{2}$, respectively. 
 $a_{\vect{k},\sigma}^\dagger$ creates a photon with polarisation $\sigma' = \pm 1$.
$p_{cv}$ is  the optical or momentum matrix element 
between valence and conduction band.
The factor $\gamma$ in equation (\ref{eq:h}) accounts for the overlap
of electron and hole spin states
\cite{ bassani75}; if spin-orbit coupling is neglected then
this becomes simply a factor of two \cite{andreani95}. We use the values
from Ref. \onlinecite{andreanibassani}.

The luminescence is determined by decay from all possible bound states in the semiconductors into
other bound states through the emission of a photon. The rate of generating a photon per unit 
photon energy corresponds to the expectation value of the number
of photons after a time $t$:
\begin{align}\label{eq:photon}
W (\hbar \omega) = \sum_{\vect{k}, \sigma} \frac{\lal a_{\vect{k},\sigma}^\dagger(t) a_{\vect{k},\sigma}(t)\rl}{t} 
\delta(\hbar \omega - E(\hbar \omega,\sigma))\textrm{,}
\end{align}
where $\omega$ is the frequency corresponding the the emitted photon with wavevector $\vect{k}$,
and $E(\hbar \omega, \sigma)$ is the dispersion of the photon in the medium. 
In this article, we consider luminescence from a biexciton in its ground state as a first approximation 
to understand luminescence from bound states in semiconductors; the result will hopefully allow us to estimate 
to what extent bound states will need to be considered,
 or whether the consideration of individual excitons 
suffices.

In order to obtain a more explicit equation for the luminescence than equation 
(\ref{eq:photon}), we first of all need to discuss the wavefunction of our biexcitonic
 ground state with respect
to which the expectation value of equation (\ref{eq:photon}) is calculated.
Previous calculations of luminescence from biexcitons have used estimates or
simplified versions for the biexciton wavefunction \cite{Ivanov};
In contrast,
the biexciton wave function $\Psi_{\rm BE}$ we use is of the form
\begin{equation}
\Psi_{\rm BE}({\bf R}) = \exp\left[J({\bf R})\right] \Psi_S({\bf R})
\;,
\end{equation}
where $J({\bf R})$ is a Jastrow factor of the Drummond-Towler-Needs
form\cite{drummond_jastrow_2004} containing only two-particle correlations, and
\begin{align*}
\Psi_S({\bf R}) &=
\phi_1(r_{13}) \phi_1(r_{24}) \phi_2(r_{14}) \phi_2(r_{23}) \\ &+ 
\phi_2(r_{13}) \phi_2(r_{24}) \phi_1(r_{14}) \phi_1(r_{23}) \;,
\end{align*}
where $r_{ij}$ is the distance between particles $i$ and $j$, with
electrons being particles 1 and 2, and holes being particles 3 and 4.
The pairing orbitals $\phi_n(r)$ are of the form
\begin{equation}
\phi_n(r) =
\exp\left[
  \frac {-r^2} {p_{1,n}(p_{2,n}+r)} +
  \frac {\Gamma p_{2,n} r} {p_{2,n}+r} \right]
 \;,
\end{equation}
where $p_{1,n}$ and $p_{2,n}$ are optimisable parameters and
 $\Gamma$ is a constant such that the Kato cusp conditions
\cite{kato_cusp_1957} at electron-hole coalescence points is
reproduced by $\Psi_S$; electron-electron and hole-hole cusps are
introduced via the Jastrow factor.

Given a trial wave function $\Psi$, assumed to be real for simplicity, the
variational Monte Carlo (VMC) method is capable of evaluating the variational
estimate $E_\Psi$ of the ground-state energy $E_0$,
\begin{equation}
E_\Psi = \frac
         {\int \Psi({\bf R}) {\hat H}({\bf R}) \Psi({\bf R})
               {\rm d}{\bf R}}
         {\int \left|\Psi({\bf R})\right|^2 {\rm d}{\bf R}}
\geq E_0
\;,
\end{equation}
by evaluating the local energy,
\begin{equation}
E_L({\bf R}) = \frac {{\hat H}({\bf R}) \Psi({\bf R})} {\Psi({\bf R})}
\;,
\end{equation}
at a set of $M$ configurations $\{{\bf R}_i\}$ distributed according to
the square of the trial wave function $\left|\Psi({\bf R})\right|^2$,
and evaluating the average
\begin{equation}
E_\Psi \approx E_{\rm V} = \frac{1}{M} \sum_{i=1}^M E_L({\bf R}_i)
\;,
\end{equation}
where the ``approximately equal'' sign refers to the statistical error
due to the finite size of the sample.  In addition, the VMC method 
optimises the parameters in the trial wave function by
means of minimizing $E_{\rm V}$ with respect to the parameters, using
techniques such as the modified linear least-squares method developed by
Umrigar \cite{umrigar_emin_2007}.

The above trial wave function is then optimised within VMC for different
mass ratios using the \textsc{casino} code\cite{needs_casino_2010}.  The
VMC energies are presented in Table~\ref{table:vmc_results}, where we
also report the variance of the local energies $\sigma_{\rm V}^2$,
which is an additional measure of the quality of a wave function.  The
variational energies obtained at all mass ratios contain a large
fraction of the exact binding energy, indicating an accurate description
of these systems. 
A comparison of VMC wavefunctions 
with both experimental and theoretical values was performed in \cite{leedn09}
for quantum well structures. 

\begin{table}[!ht]
\begin{center}
\begin{tabular}{r@{.}lr@{.}lr@{.}l}
\hline
\multicolumn{2}{c}{$m_h/m_e$} &
\multicolumn{2}{c}{$E_{\rm V}$} &
\multicolumn{2}{c}{$\sigma_{\rm V}^2$} \\
\hline \hline
 1837&36222 & -1&1502(4)  & 0&131(1) \\
  183&73622 & -1&1335(2)  & 0&01831(5) \\
   18&37362 & -1&0338(1)  & 0&01587(5) \\
    1&83736 & -0&66725(8) & 0&00511(2) \\
\hline
\end{tabular}
\caption{\label{table:vmc_results} VMC energies and variances of the
local energy for different hole-to-electron mass ratios.}
\end{center}
\end{table}

Using the biexcitonic wavefunction calculation with VMC
as our ground state, we can
now transform equation (\ref{eq:photon}) into Fermi's golden rule for the transition 
rate $W(\hbar \omega)$ per energy range of a particular photon \cite{Chuang}. 
Fermi's golden rule corresponds to a perturbative treatment of the interaction
\cite{Schiff}, and has already been used by Elliot \cite{elliott57} for a calculation
of exciton luminescence. 
The transition rate per energy range requires 
 a sum over all final states of both exciton and photon,
\begin{align}\label{eq:fermisgolden}
W(\hbar \omega) & = \frac{2\pi}{\hbar}\sum_{\vect{k},\vect{K}, \{n_i\}} | 
    \lal \Psi_{\rm{E,\{n_i\}},\vect{K}} | H' | \Psi_{\rm{BE}} \rl|^{2} \\
   & \qquad \delta (\epsilon_i -\epsilon_f(\vect{k},\{n_i\})-\hbar \omega)
    \delta (\hbar \omega - \hbar c |\vect{k}|) \textrm{.} \nonumber
\end{align}
As final exciton momenta $\vect{K}$ are constrained by momentum conservation, 
 the corresponding sum disappears, and the momentum of the surviving exciton 
is $-\vect{k}$.
The final excitonic state with wavefunction
$\Psi_{\rm{E,\{n_i\}},\vect{K}}$  is labelled by the centre of mass  $\vect{K}$ and
 the various eigenstates of the exciton $\{n_i\}$, which are standard 
hydrogenic solutions. The quantum numbers  
$\{n_i\}$ correspond to $n$, $l$, and $m$ for the bound and $k$, $l$, and $m$ 
for the continuum states. As the biexciton is in its ground state, we 
obtain from angular momentum conservation that $l = m = 0$ for the surviving exciton.
$\epsilon_i$ is the ground state biexciton energy, and $\epsilon_f (\vect{k},n)$
the energy of the final state.

We now insert 
$H'$ from equation (\ref{eq:h}) into equation (\ref{eq:fermisgolden})
and notice that it is necessary to perform several real space integrals in
order to simplify the matrix element.
After a separation of centre of mass and relative coordinates, the relative exciton
wavefunction $\phi_{\rm{E,n}}$ can be expressed in terms of a single coordinate 
$\vect{r}_1 = \vect{r}_{e1} - \vect{r}_{h1}$, where $\vect{r}_{e1}$ and $\vect{r}_{h1}$
refer to the coordinates of electron and hole of the surviving exciton.
The relative biexciton wavefunction  $\phi_{\rm{BE}}$
depends on this coordinate $\vect{r}_1$,
and the coordinates of the second electron and hole, $\vect{r}_{e2}$ and $\vect{r}_{h2}$.
In our relative coordinate system these are represented by a vector to their centre of mass,
$\vect{r}_2 = (\vect{r}_{e1}+\vect{r}_{h1} - (\vect{r}_{e2} + \vect{r}_{h2}))/2$, and 
a relative coordinate $\vect{r}_3 = \vect{r}_{e2} - \vect{r}_{h2}$.
The application of the dipole approximation, $\vect{p}_{{\rm op}} \approx 0$
requires the annihilating electron and hole to be at the same point in space. 
Thus, this relative coordinate is zero, and so
 $\phi_{\rm{BE}}$ has the functional dependence 
 $\phi_{\rm{BE}}(\vect{r}_1,\vect{r}_2,\vect{0})$.
For the bulk system, the centre of mass coordinates can be integrated over analytically.
We then obtain a formula for the transition rates $R$ into the bound states $n$,
\begin{align} \label{eq:w}
R_n &=
    \frac{p_{cv}^2e^2}{m_{0}^2} 
       \frac{\sqrt{\varepsilon}}{2 \varepsilon_0 \pi}\frac{\omega_n}{\hbar c^3}
           \\
   &\quad 2
                 \Big (\int d \vect{r}_1 d \vect{r}_2  \phi_{\rm{E,n}}(\vect{r}_1)^{*} 
                 \phi_{\rm{BE}}(\vect{r}_1,\vect{r}_2,\vect{0})  
                 f(\vect{r}_1, \vect{r}_2)
                  \Big )^2   \textrm{.} \nonumber
\end{align}
where we have introduced the function $f(\vect{r}_1, \vect{r}_2)$ to model confinement in 
a quantum dot. We will discuss this function in section III; for bulk, $f(\vect{r}_1, \vect{r}_2) =1$. The additional factor of 2 in equation (\ref{eq:w}) comes from the
fact that both excitons in the biexciton can recombine.

The frequency $\omega_n$ in equation (\ref{eq:w}) is
the frequency associated with a specific transition. We neglect its 
momentum-dependence: The frequencies that contribute most to the emission spectrum
 will be close to the gap energy $E_g$ and hence
we approximate $\omega_n \to E_g/\hbar$ from here onwards. Our results confirm later that
this treatment is indeed adequate. Thus the rate $R_n(n)$ in equation (\ref{eq:w})
is a function of quantum number $n$ only.

Since $R_n$ is the transition rate into a particular bound state,
$R_n$ and $W(\hbar \omega)$ are related by
\begin{align}\label{eq:allw}
W(\hbar \omega ) = \sum_n R_n \delta(\hbar \omega - \hbar \omega_n) + W_{\rm{continuum}} (\hbar \omega)\textrm{,}
\end{align}
where $ W_{\rm{continuum}} (\hbar \omega)$ describes emission into 
continuum states. 
$ W_{\rm{continuum}} (\hbar \omega)$ can be obtained analagously to 
equation (\ref{eq:w}), with the only difference that 
the sum over discrete quantum numbers $n$ in
equation (\ref{eq:fermisgolden}) becomes an integral over the continuum states
and the continuum exciton eigenfunctions
are confluent hypergeometric functions.

\subsection{Results \label{sec:s1r}}
\begin{figure}
 \includegraphics[width = 0.44\textwidth]{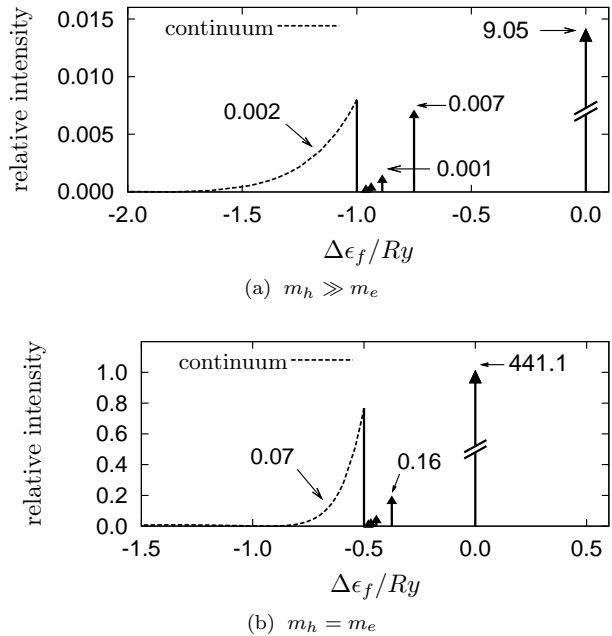}
\caption{Bulk emission spectra (overlap element as shown in Table
 \ref{tab:overlap}
 versus the energy difference of the surviving exciton with respect to its
ground state, $\Delta \epsilon_f$) for the two limiting mass ratios. 
The emitted photon energy corresponds to
$\hbar \omega = E_g- B_{2X} - 1Ry + \Delta \epsilon$ for $m_h$ $\gg m_e$ and
$\hbar \omega = E_g- B_{2X} - 0.5Ry + \Delta \epsilon$ for $m_h = m_e$,
where $B_{2X}$ is the biexciton binding and a single exciton has 
energy $E_g -1Ry$ and $E_g - 0.5Ry$ for these two mass ratios, respectively.  
Note that the peaks from the discrete bound states are delta functions and hence represented by arrows,
labelled by their intensity. 
While it is clear that the continuous part of the spectrum decays to zero for
large energies $E$, there is a possibility for a small but finite overlap for intermediate $E$. The number labelling the continuum is its integrated weight. \label{fig:spectra}}
\end{figure}
\begin{table}
\begin{center}
\caption{Dimensionless overlap elements,
$2 \int d \vect{r}_1 d \vect{r}_2  \phi_{\rm{E,n}}(\vect{r}_1)^{*} 
 \phi_{\rm{BE}}(\vect{r}_1,\vect{r}_2,\vect{0})$, of excitonic bound  (quantum number $n$)
and continuum states with the biexciton wavefunction for various mass ratios.
 The mass ratios ($m_h$:$m_e$)
correspond to decreasing the $m_h$ from 1837$m_e$ (in hydrogen) by factors of 10.
Individual elements are converged to approximately \%0.5 accuracy. An estimate of the absolute
accuracy including the uncertainty of the wavefunction cannot be given,
 but we tentatively estimate ~10\% accuracy from comparisons with a different variational form.
`Total' indicates the sum over the calculated bound and continuum states, and the
result from  equation (\ref{eq:T}) is shown in the
last row.  
We note that for $n \ge 2$ the values drastically decrease compared to $n=1$ (base
change as shown in the first column); however, these higher $n$ 
 matrix elements depend very sensitively on the form of the wavefunction.
 \label{tab:overlap}}
\begin{tabular}{l |  r@{.}l r@{.}l r@{.}l r@{.}l r@{.}l }
\hline
$m_h$ & \multicolumn{2}{c}{1837}  & \multicolumn{2}{c}{183.7} & \multicolumn{2}{c}{18.3} &\multicolumn{2}{c}{1.8}&\multicolumn{2}{c}{1.0}  \\
\hline
\hline
$n=$1              &  9&1 & 20&4 & 79&6 & 427&0 & 441&1\\
$n=$2 (/$10^{-3}$) &  6&6 &  8&0 &  0&1 & 305&2 & 161&5 \\
$n=$3 (/$10^{-3}$) &  1&0 &  1&0 &  0&4 &  18&3 &  34&8 \\
$n=$4 (/$10^{-3}$) &  0&3 &  0&3 &  0&2 &   4&8  & 13&9 \\
$n=$5 (/$10^{-3}$) &  0&2 &  0&1 &  0&1 &   4&1  &  6&8 \\
bound total        &  9&1 & 20&4 & 79&6 & 427&3  &441&3\\
\hline
continuum (/$ 10^{-2}$) &  0&3 &  0&4 &  1&0 &  5&4&  7&3 \\
\hline
total &  9&1 & 20&4 & 79&6 & 427&3 & 441&4 \\ 
Equation (\ref{eq:T}) & 9&1 & 20&4 & 79&6 & 427&4 &  441&5\\
\hline
\end{tabular}
\end{center}
\end{table}
The emission spectra from bulk for the two limiting mass ratios,
$m_h = 1837m_e$ (corresponding to hydrogen), and 
$m_e = m_h$ are shown in Figure \ref{fig:spectra}.
As we have not included broadening, the delta-function peaks for transitions into the bound states
are indicated by arrows, and labelled with the corresponding delta-function coefficient.
Figure  \ref{fig:spectra} shows clearly that the transition
into the ground state dominates. The spectra for the intermediate mass ratios show a similar 
behaviour. 

Individual overlap integrals for all different mass ratios are shown in Table 
\ref{tab:overlap},
which also contains the results obtained from equation (\ref{eq:T}).
The discrepancy between the results from the individual overlaps (`total')
and from equation (\ref{eq:T}) is a rough estimate of the numerical errors of the 
calculation. The agreement of these values shows that the 
results for the ground state are numerically stable. 
 Further error associated with the results are due to the
approximate shape and the parameters involved in the wavefunction.
 A different variational form
for hydrogen gave bound state transition rates within 10\% 
and continuum transition rates within about 30\% of those in Table \ref{tab:overlap}.
While one can tell that the overlaps with the higher excited states
decrease with $n$, the actual individual numbers depend very sensitively on detailed
properties of the numerically determined wavefunction and hence should
not be interpreted too closely. 

The general shape of the  continuous part of our emission spectrum should nevertheless
be correct. The shape is notably different to the shape of the single exciton absorption
spectrum calculated by Elliott\cite{elliott57}. This single exciton absorption spectrum
should correspond to a single exciton emission spectrum, provided the single exciton emission
is possibe: a single exciton in free space cannot recombine due to momentum conservation,
but recombination is possible when impurities are present and momentum conservation is relaxed.
Nevertheless, the continuum part of the single exciton and biexciton emission spectra are different. Similar to single exciton absorption, the biexciton continuum has a finite onset,
but falls off rapidly with $E$,
 while the exciton absorption continuum line increases
 square-root-like for large energies.
We expect the general trend of our spectrum to be correct, although 
we note that our continuum curve at energies of order $E \approx 1 Ry$
 is numerically sensitive to details of the Monte Carlo wavefunction.
 In this energy range we sometimes obtain a small but finite overlap,
which is due to the numerical integration being very sensitive
to the variational form of the wavefunction in this energy range. 

The continuum emission decreases with $E$ due to the decreasing wavefunction overlap of
the initial state with the higher energy exciton state after photon emission.
The fast decrease of the continuum emission 
 in Figure \ref{fig:spectra} is especially notable for equal masses. 
Such a biexciton involving particles with equal masses is more loosely bound than
one with a heavier hole. Loosely bound excitons have a smaller binding 
energy and thus less energy is available from  the recombination.
As the binding energy decreases the recoil momentum
 of the surviving exciton, $\vect{K} =-\vect{p}_{op} $,
decreases towards zero, where recombination is prohibited by momentum conservation.
The weaker fall-off of the continuum curve for the tighter bound biexciton
suggests that non-radiative processes, where the surviving exciton absorbs all the
energy available from the annihilation, are more likely for heavier holes.
When non-radiative decay is the main recombination mechanism, exciton lifetimes
are dependent on charge carrier density or temperature \cite{haug78, beattiel59}.

The most notable feature of our calculation is the predominance of the $1s$ peak.
This predominance implies that the surviving exciton is not affected by the
recombination of the other-electron hole pair, even though we initially assume a strongly
 interacting bound state. Our model thus justifies treating excitons as non-interacting
for the purpose of optical recombination. However, we speculate that interactions
become more important if higher excited initial states are present.
 
The lineshapes in emission spectra from biexcitons 
are often broadened due to scattering with phonons \cite{akiyamakmk90,nagaikg02}.
This broadening smears out the different 
peaks from a variety of possible final states, and hence no direct comparison 
with our spectra is possible.
 Similar problems prevent measurements of lifetimes in bulk \cite{herzp02}.
In addition, higher excited initial biexciton states may be present.
However, our tentative spectra for bulk can be used as a reference point for other systems.

\section{Lifetimes}
We now turn to the calculation of the biexcitonic lifetimes. We first discuss how to
include polariton effects and present a general formula for the biexcitonic lifetime
before turning to a brief discussion of lifetimes for a confined quantum dot. The results
are also divided into a section on bulk and quantum dot systems. We compare to experimental
and theoretical values in each of these subsections.

\subsection{Methods \label{sec:s2m}}
The overall rate for biexciton decay, which corresponds to the inverse biexcitonic
lifetime, is made up from the
 individual transition rates in equation (\ref{eq:allw}), 
\begin{align} \label{eq:lifetimetot2}
\frac{1}{T} &= \int d(\hbar \omega) W (\hbar \omega) \nonumber \\
  & = \sum_{n} R_n 
   + \int d(\hbar \omega) W_{\rm{continuum}} (\hbar \omega) \textrm{.}
\end{align}

We note that the sum over excitonic states
 in equation (\ref{eq:lifetimetot2})
is complete and hence can be understood as a resolution of identity, 
and thus total biexciton decay rate can also be expressed in terms of
 the biexciton wavefunction only,
\begin{align} \label{eq:T}
\frac{1}{T}& = 
   \frac{p_{cv}^2e^2}{m_{0}^2} 
     \frac{\sqrt{\varepsilon}}{2 \varepsilon_0 \pi} \frac{E_g}{\hbar^2 c^3}
       \nonumber \\
&    \quad 2 
     \int   d\vect{r}_{2}' d \vect{r}_1 d\vect{r}_{2} 
        \phi_{\rm{BE}}(\vect{r}_1,\vect{r}_2, \vect{0})
        \phi_{\rm{BE}}(\vect{r}_1,\vect{r}_{2}',\vect{0})
     \textrm{.}
\end{align}
This expression should provide a check for the numerical errors associated with the integral.

The photon generation rate in equation (\ref{eq:photon}) does not yet include the polaritonic
eigenstates. We mention earlier that these are important for nearly all semiconductors, 
since the biexciton binding energy $\eps_m \ll \Omega_c$ (the polariton splitting parameter)
in almost all materials. 
The exact calculation of
 lifetime from a semiconductor would thus require the rate of polariton generation per unit energy
\begin{align}\label{eq:polariton}
W (\hbar \omega) &= \frac{1}{2} \sum_{\vect{k}, s, \sigma} 
\frac{\lal \xi_{\vect{k},s, \sigma}^\dagger(t) \xi_{\vect{k},s, \sigma}(t) \rl}{t}\nonumber \\ 
& \qquad \qquad \delta(\hbar \omega - E_{\textrm{polariton}}(\vect{k}, s, \sigma))\textrm{,}
\end{align}
where $\xi_{\vect{k},s, \sigma}^\dagger(t)$ creates a polariton with momentum $\vect{k}$ and spin $\sigma$ in state $s$
and $E_{\textrm{polariton}}$ is the corresponding dispersion. The factor of one half
avoids double-counting, since in the biexciton decay two polaritons are generated.
\footnote{
Equation (\ref{eq:polariton}) calculates the general decay rate into a two-polariton state and not 
the rate at which photons are emitted from the semiconductor, since it is this general decay rate
 that is experimentally accessible. }
 In general the photon can be reabsorbed to form another biexciton
and reemitted in many cycles. This would require a self-consistent treatment such as performed
in Ref. \onlinecite{Ivanov}, which we expect to be important for materials with
large $\Omega_c/\eps_m$.
 As we saw previously that the decay rate is dominated by the decay rate into the 1s exciton,
 we assume that the created polariton is a superposition of a 1s exciton and a photon and do not
consider excited exciton states.
 We also use the resonant approximation for the
polaritonic eigenstates, which we expect to cause an error of no more 
than 15\%. With these approximations, our expression for the inverse lifetime becomes
\begin{align}\label{eq:tp}
\frac{1}{T} &= \frac{\gamma}{2} \times
| \lal \Psi_{\rm{E,\{n_i\}},\vect{K}} | \Psi_{\rm{BE}} \rl|^{2}  \times \Big\{ 2 (E_0-2E_g)^2 +\nonumber \\
&\frac{[E_0 (-E_0 +2 E_g)+ \Omega_{c}^2]^2 \Omega_{c}^2}{2(E_0 -2 E_g)^2 [(E_0-2E_g)^2+ \Omega_{c}^2]}  \Big\} \nonumber \\
 & \times \frac{p_{cv}^2 e^2}{m_{0}^2} \frac{\sqrt{\eps^3} }{\pi \eps \eps_0 E_g/\hbar} \frac{1}{(\hbar c)^3}\textrm{.}
\end{align}

Because lifetimes are longer and hence more extensively studied for confined systems,
we wish to study the effect 
 of confinement on the recombinative lifetime,  and
we consider a 3D quantum dot system. 
We assume a spherical quantum dot of radius $d$, where the photon wavelength
$\lambda \gg d \gg a_E, a_B$, where $a_E$ and $a_B$ are relevant exciton 
and biexciton lengthscales. The exciton lengthscale $a_E$ is the distance between
the electron and hole in the surviving exciton, and the biexciton lengthscale
$a_B$ is the distance between the centres of mass of the two excitons in the 
biexciton. Such a dot is large enough to only affect the 
centre of mass, but not the relative wavefunction.
The centre of mass wavefunctions can then be expressed in terms of spherical 
Bessel functions, and we obtain the constraint that the centre of mass wavefunction
 of the surviving exciton must also be in the ground state. 
Again, we use the 1s exciton-polariton wavefunction;
the corresponding lifetime is $R(n=1)$ in equation (\ref{eq:w}).
The effect of confinement is reflected in the function $f(\vect{r}_1,\vect{r}_2)$
in equation (\ref{eq:w}), where 
\begin{align*}
f(\vect{r}_1,\vect{r}_2) = \frac{4d}{\pi } \frac{ \sin (\frac{\pi}{d} \frac{|\vect{r}_2|}{2})}{|\vect{r}_2|} \textrm{.}
\end{align*}

\subsection{Results \label{sec:s2r}}
We have calculated lifetime estimates for a number of materials, and compare these
to experimental and other theoretical lifetime estimates in this section. 
The parameters required for the calculations of these lifetimes for CuCl, GaAs, ZnSe and InGaAs are given
in table \ref{tab:table}, and table \ref{tab:table2} contains quantum dot values for
 those parameters which differ between bulk and quantum dot systems. Since 
the calculated lifetimes depend quite
sensitively on the correct parameter and there is some variation
of parameters in the literature, this comparison is quite difficult. One example is e.g. the
InGaAs mass ratio, where different groups use bulk or quantum well or interpolated quantum well mass
ratios, or the experimental GaAs lifetime. Here
  we quote an experimental value for GaAs bulk lifetime of 1.8\,ns in table \ref{tab:lifetimes} merely for completeness, since we expect the real lifetime to be around 10-100\,ps \cite{jfprivate}. 
 This estimate is based on the fact that
the bulk biexciton lifetime should be smaller than the lifetime in confined system like quantum
dots or quantum wells, where the biexciton binding energy is higher \cite{phillipslovering}. 
 For quantum wells, the biexcitonic lifetime is expected to be around half the excitonic lifetime \cite{spiegel, charbonneau},
which in Ref. \onlinecite{charbonneau} is quoted to be of order 200\,ps, while Ref.  \onlinecite{devaud}
contains an estimate of 10\,ps for the excitonic lifetime and
Ivanov \textit{et al.} \onlinecite{Ivanov}. quote lifetimes of order of magnitude 1-10\,ps.
There exists nevertheless a higher lifetime estimate 
 of 1\,ns \cite{cingolani} for quantum dots; 
however Ref \onlinecite{devaud} cite potential impurities as a reason for lifetimes of this order of magnitude in
quantum wells.  
Thus, the variety in experimental values makes
the comparison sometimes somewhat difficult.

\begin{table*}
\caption{Semiconductor parameters with references. $\eps_{m}$ is the biexciton binding energy. \label{tab:table}}
\begin{tabular}{| c | c c c c |}
\hline
constant & GaAs & ZnSe & CuCl &  $\rm{In}_{0.6} \rm{Ga}_{0.4}$ As \\
\hline
\hline
$\eps_{m}/ meV$ &0.13\cite{kuther98}, 0.45\cite{kuroda06},  0.7$\pm$ 0.2\cite{steiner89} &
                 3.5\cite{yamada95binding} &
 32 \cite{Ueta} , 34 \cite{Ivanov},  42 \cite{itoh1989nonlinear}, 42\cite{Park00}
              & 2\cite{narvaezGBZ05}\\
$E_g$         & 1.5\cite{Chuang} & 2.8\cite{nozue} & 3.2\cite{Ivanov}  & 0.7\cite{nozue}\\
$\eps$           & 13.1 \cite{Chuang}& 8.1 \cite{nozue} & 5.6\cite{Ivanov}, 5.0 \cite{AndreanidelSole}  & 14.3 \cite{Singh}, 14.03 \cite{stiergm99}\\
$m_h/m_0 $       & 0.5 \cite{Chuang} & 1.7\cite{nozue} & 1.8 \cite{Ueta}, 2.0 \cite{AndreanidelSole} 
                 &   0.46\cite{Adachi}\\
$m_e/m_0 $       & 0.067 \cite{Chuang} & 0.017\cite{nozue}&  0.5 \cite{Ueta, AndreanidelSole} & 0.04 \cite{stiergm99}, 0.05\cite{liao11}, 0.067\cite{wimmer06} \footnote{from bulk values, but inclusion of strain effects in Hamiltonian} \\
$\Omega_c /meV$  & 15.6 \cite{Ivanov}  & 100\cite{nozue}  & 191 \cite{Ivanov} & 7 \\
$E_p$  \footnote{$E_p$ and $\Omega_c$ are related via the formula 
$\Omega_c = 2 \sqrt{\gamma} \sqrt{2\pi} p_{cv} \hbar/(m_0 \sqrt{4 \pi \epsilon \omega_0 \hbar} \sqrt{\pi a_{E}^3})$ and thus one can be derived from the other. The value of $E_p$ in InGaAs where we do not quote a source was generated by this formula; note that the $\Omega_c$ of 100$\mu eV$ measured by \cite{laucht} is for a quantum dot in a nanocavity and thus not immediately applicable. For all other materials $E_p$ is taken from the quoted source.}          &  25.7 \cite{Chuang} & 29.56 (from \cite{willatzen}) & 2.3  & 21.7 \cite{stiergm99} \\
\hline
\end{tabular}
\end{table*}

\begin{table*}
\caption{Quantum parameters with references. $\eps_{m}$ is the biexciton binding energy. \label{tab:table2}}
\begin{tabular}{| c | c c |}
\hline
constant & CuCl  & $\rm{In}_{0.6} \rm{Ga}_{0.4}$ As \\
\hline
\hline
$\eps_{m} /meV$  &50\cite{NairBeCuCl} (3nm), 51\cite{Park00}, 60\cite{masumotoBeCuCl} (3nm) & 2\cite{narvaezGBZ05} \\
$m_h/m_0 $      & 1.8 \footnotemark[1] \footnotetext[1]{from bulk} & 0.125\cite{bulaevloss05}\footnote[2]{from interpolation between
InAs and GaAs masses, using quantum well masses from \cite{chan86} and \cite{wimbauer94}},
0.2 \cite{liao11}, 0.46\footnotemark[1], 0.5\cite{wimmer06} \footnote[3]{from bulk values, but inclusion of strain effects in Hamiltonian}\\
\hline
\end{tabular}
\end{table*}

\begin{table}
\caption{Lifetime estimates for several semiconducting materials and experimental values for comparison. $\tau_{\rm theor}$ 
corresponds to other theoretical calculations of biexciton lifetimes that are discussed in the text.\label{tab:lifetimes}. We expect the correct lifetime for GaAs to be of order 10-100\,ps (see main text for discussion).}
\begin{tabular}{ c  c c c | c c }
\hline
   & \multicolumn{3}{c|}{bulk} & \multicolumn{2}{c}{quantum dot} \\
   & GaAs & ZnSe & CuCl & CuCl & $\rm{In}_{0.6} \rm{Ga}_{0.4}$ As \\
\hline
\hline
 $\tau /ps$ &  0.5 -2.7  
 & 0.4 - 2.0 & 8.8-47.3 &  47.4-66.9 & 5.5-12.8 \\
 $\tau_{\rm exp}/ps$        &    1800 \cite{steiner89} &   40\cite{yamada95lifetime} & 18-27\cite{hasuoCuCl, IvanovCuClexp} & 65\cite{edamatsu96} & 500 \cite{ulrichf05} \\
$\tau_{\rm theor}/ps$  & - & - & 24 \cite{Ivanov} & - & 500 \cite{narvaezGBZ05, wimmer06}\\
\hline
\end{tabular}
\end{table}

\subsubsection{Bulk}
We now discuss polariton lifetimes for bulk materials and the variation of the lifetime with mass
ratio.

Without including polariton effects the lifetime for CuCl according to equation \ref{eq:tp} would
be 20 and 100\,ps for mass ratios 1.8 and 18, respectively. The experimental lifetimes are
between 18 and 24\,ps \cite{hasuoCuCl,IvanovCuClexp} for a mass ratio $m_h/m_e \approx 5$, 
which shows that the inclusion of polaritonic states (giving instead lifetimes 9\,ps and 47\,ps for 
the same mass ratios) is indeed crucial.
For the three semiconductors GaAs, ZnSe and CuCl experimental and our
calculated lifetimes are shown in table \ref{tab:lifetimes}. We expect our method to work well for
small ratios $\Omega_c/\eps_{m}$, while for large $\Omega_c/\eps_{m}$ a full bipolariton wavefunction
and thus the inclusion of recursive exciton creation and annihilation processes is necessary. 
Additional errors in our estimate are associated with intrinsic errors of the wavefunction, 
the numerical integration, the uncertainty of the material parameters, the different
geometry of the quantum dot and the fact that experimentally, the exciton might leak out.

Indeed, our lifetime estimates for CuCl between 9 
and 47\,ps for mass ratios 1.8 and 18,
respectively, agree reasonably with the experimental value for the intermediate mass ratio of
$m_h/m_e \approx 5$. Ivanov et. al. \cite{Ivanov}
also obtain excellent agreement when using the bipolariton wavefunction, but quote a poor lifetime
estimate when using the giant oscillator model \cite{hanamura} (4\,ps). The improved wavefunction and a different
weight in front of the process responsible for creating two lower state polaritons in equation \ref{eq:tp}
account for this difference.

For ZnSe and GaAs the increased ratio $\Omega_c/\eps_m$ indicates 
that polariton effects should matter 
more. Indeed, our lifetime estimates are lower than the experimental 
values by about factor of 10.

\begin{figure}
   \includegraphics[width = 0.46 \textwidth]{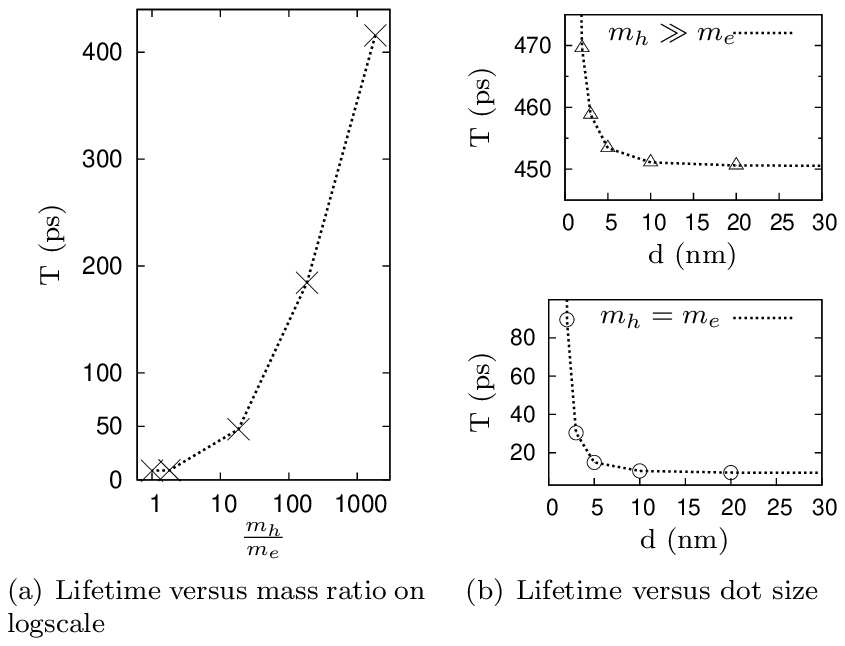}
\caption{Left panel: 
Biexciton lifetimes (in ns) in 3D bulk, calculated with approximate values for a CuCl
semiconductor, for different mass ratios as shown in Table \ref{tab:overlap}.
Right panel: Biexciton lifetime for hole masses $m_h = 1837m_e$ and $m_h = m_e$ 
for different CuCl quantum dot radii. \label{fig:life}}
\end{figure}

Figure \ref{fig:life} shows plots of the dependence of the lifetime on mass ratio, 
with parameters for 
CuCl.
The decrease of lifetimes on approaching the equal mass ratio limit in  
Figure \ref{fig:life}(a) is initially
counterintuitive as electron and hole are much more likely to be at the same place
for annihilation in the more tightly bound hydrogenic exciton.
We find however that the probability for recombination is highest when the
two excitons within the biexciton are far apart.
This dependence on the mass ratio can be explained by the relative importance 
of the biexciton lengthscale $a_B$ (the distance between the centre of 
masses of the two excitons) and the exciton lengthscale $a_E$ (the
distance between electron and hole within an exciton) in the overlap integral in equation 
(\ref{eq:w}).
\footnote{We note that the lengthscales $a_B$ and $a_E$ are only appropriate for describing this problem
if the biexciton can be envisaged as being made up of two excitons that retain their identity
in the bound state. This picture of a biexciton has been found to be correct at least 
 in two dimensional systems with
 large electron-hole layer separation  \cite{leedn09,szimmermann08, mfogler08}.}
 The reason for the influence
of the distance $a_B$ 
on the recombination rate comes from the overlap integral in equation 
(\ref{eq:w}): 
the single exciton final state is delocalised. 
The overlap with the biexciton wavefunction is thus higher for a biexciton
wavefunction in which the centres of mass of the excitons are more delocalised, i.e.
where $a_B$ is large.
The lengthscale ratio $a_B/a_E$ 
is inversely proportional to the ratio of biexcitonic to excitonic binding energies,
and the latter decreases towards the equal mass ratio limit for
both two and three dimensional systems \cite{banyaigeh87, xie01}; thus, 
in the equal mass biexciton, the excitons are loosely bound and are further apart.

\subsubsection{Quantum dot}
We now turn to lifetimes of confined systems, and discuss both the impact of confinement
on the lifetime as well as some calculated lifetimes for two materials.

Figure \ref{fig:life}(b)
shows the effect of confinement in a spherical quantum dot where confinement principally
affects the centre of mass wavefunction. 
Stronger confinement increases the exciton localisation
and thus the increase of lifetime with confinement shown in Figure \ref{fig:life}(b)
is consistent with the previous argument of increased annihilation probability for
biexcitons made up of a loosely bound exciton pair.

 Unfortunately, experiments are not typically in the regime where confinement principally
affects the centre of mass wavefunction, and so accurate numerical predictions of
the biexciton lifetimes in these cases require specific
calculations for the particular structures. Such specific calculations have for example been 
performed for In$_{0.6}$Ga$_{0.4}$As quantum dots.
Using a configuration interaction method,
Narvaez \textit{et al.} [\onlinecite{narvaezGBZ05}] obtain the exact experimental value
of 0.5\,ns for a 
252$\times$252$\times$75\,$\rm{\AA}$
In$_{0.6}$Ga$_{0.4}$As 
quantum dot, which agrees well with the experimental lifetime of 0.5\,ns of a
 dot with height 20$\AA$ and a square base with length 150-200$\AA$\cite{ulrichf05}.
  Wimmer \textit{et al.} [\onlinecite{wimmer06}] perform a Monte Carlo
optimisation of wavefunctions in anisotropic quantum dots for three materials, among
them In$_{0.6}$Ga$_{0.4}$As, and obtain excellent lifetime estimates for all three materials.
 Since the exciton Bohr radius for this system is about 200$\AA$
and the wavefunction is thus strongly affected by confinement effects, our wavefunction
is  a inaccurate estimate for this system. Additionally, our calculation does not take
into account strain effects and anisotropy of the hole masses,
which are nonnegligible effects for InGaAs. For example, it is well known that these effects 
have a large impact on gyromagnetic ratio \cite{nakaoka}, where realistic k.p simulations specific to the material
 provide estimates adequate to the experimental data \cite{jovanov}.  Neglecting these effects and
approximating this quantum dot by a spherical quantum dot of radius 100$\AA$, we obtain a lifetime estimate of 6-13\,ps, 
which is clearly inaccurate. 

For 3\,nm CuCl quantum dots, the 
quantum dot radius is larger than the excitonic Bohr radius of 6.8$\AA$ and thus
our wavefunction should still be viable.
Here, obtain a biexciton lifetime estimate of 47 -67\,ps for our two
limiting mass ratios, which is
 comparable to the experimental value of 65\,ps. This shows that our method is indeed
reasonable for the systems with small polariton effects, and can provide good estimates of the lifetime.

\vspace{0.4cm}
\section{Conclusions \label{sec:concl}}
In this article we have investigated the lifetime and emission spectrum of a biexcitonic
system in different mass limits. We employ the approach by Elliott
in order to find the transition rates into excitonic states using Fermi's golden
rule. We find that in our model, which assumes an initial biexciton in the ground state,
the annihilation of one electron and hole is most likely to result in an exciton
in its ground state. The predominance of this transition means that the two excitons which 
form the excitonic molecule can be treated as non-interacting for the purpose
of optical recombination. 
Our estimates of lifetimes for different mass ratios are comparable to experimental values.

In this calculation we have assumed a biexciton which was initially in its ground state.
The presence of higher excited states
could lead to a stronger overlap with higher excited final states,
and thus to shorter lifetimes. We have also neglected the possibility of collision with other particles
which could also lead to shorter lifetimes.
We find that the biexciton lifetime slightly increases in confined quantum dots.
Using an appropriately generated wavefunction,
our method is also applicable for smaller quantum dots or quantum wells. 

 Interesting extensions to fully model
quantum dots could include strain effects and for example the
impact of the wetting layer, for which experimental data is also available \cite{wetting}.
It would also be interesting to calculate lifetimes 
for pumped systems, where the timescale of
 the biexciton lifetimes is important 
 for an accurate estimation of the efficiency of multi-exciton
generation \cite{nozik2008multiple, witzel}.
 
 Our approach can be used as a benchmark for
comparing approximative methods for systems at finites densities. 
We also hope that the relative simplicity of the biexciton emission spectrum as found
here may imply that the development of further experimental and theoretical understanding
of emission from the dense electron-hole plasma is within reach.

\begin{acknowledgements}
M.~B. thanks R.~T.~Brierley , P.~G.~Brereton, T.~Eisfeller, J.~J.~Finley, and R.~T.~Phillips for helpful discussions, 
and the Gates Cambridge Trust for financial support. J.~K. and M.~M.~P. acknowledge funding from EPSRC Grant Nos. EP/G004714/1 and EP/H00369X/1. P.~L.~R. acknowledges funding from the EPSRC.
\end{acknowledgements}


\begin{thebibliography}{10}%
\makeatletter
\providecommand \@ifxundefined [1]{%
 \ifx #1\undefined \expandafter \@firstoftwo
 \else \expandafter \@secondoftwo
\fi
}%
\providecommand \@ifnum [1]{%
 \ifnum #1\expandafter \@firstoftwo
 \else \expandafter \@secondoftwo
\fi
}%
\providecommand \enquote [1]{``#1''}%
\providecommand \bibnamefont  [1]{#1}%
\providecommand \bibfnamefont [1]{#1}%
\providecommand \citenamefont [1]{#1}%
\providecommand\href[0]{\@sanitize\@href}%
\providecommand\@href[1]{\endgroup\@@startlink{#1}\endgroup\@@href}%
\providecommand\@@href[1]{#1\@@endlink}%
\providecommand \@sanitize [0]{\begingroup\catcode`\&12\catcode`\#12\relax}%
\@ifxundefined \pdfoutput {\@firstoftwo}{%
 \@ifnum{\z@=\pdfoutput}{\@firstoftwo}{\@secondoftwo}%
}{%
 \providecommand\@@startlink[1]{\leavevmode}%
 \providecommand\@@endlink[0]{}%
}{%
 \providecommand\@@startlink[1]{%
  \leavevmode
  \pdfstartlink
   attr{/Border[0 0 1 ]/H/I/C[0 1 1]}%
   user{/Subtype/Link/A<</Type/Action/S/URI/URI(#1)>>}%
  \relax
 }%
 \providecommand\@@endlink[0]{\pdfendlink}%
}%
\providecommand \url  [0]{\begingroup\@sanitize \@url }%
\providecommand \@url [1]{\endgroup\@href {#1}{\urlprefix}}%
\providecommand \urlprefix [0]{URL }%
\providecommand \Eprint[0]{\href }%
\@ifxundefined \urlstyle {%
  \providecommand \doi [1]{doi:\discretionary{}{}{}#1}%
}{%
  \providecommand \doi [0]{doi:\discretionary{}{}{}\begingroup
  \urlstyle{rm}\Url }%
}%
\providecommand \doibase [0]{http://dx.doi.org/}%
\providecommand \Doi[1]{\href{\doibase#1}}%
\providecommand \bibAnnote [3]{%
  \BibitemShut{#1}%
  \begin{quotation}\noindent
    \textsc{Key:}\ #2\\\textsc{Annotation:}\ #3%
  \end{quotation}%
}%
\providecommand \bibAnnoteFile [2]{%
  \IfFileExists{#2}{\bibAnnote {#1} {#2} {\input{#2}}}{}%
}%
\providecommand \typeout [0]{\immediate \write \m@ne }%
\providecommand \selectlanguage [0]{\@gobble}%
\providecommand \bibinfo [0]{\@secondoftwo}%
\providecommand \bibfield [0]{\@secondoftwo}%
\providecommand \translation [1]{[#1]}%
\providecommand \BibitemOpen[0]{}%
\providecommand \bibitemStop [0]{}%
\providecommand \bibitemNoStop [0]{.\EOS\space}%
\providecommand \EOS [0]{\spacefactor3000\relax}%
\providecommand \BibitemShut [1]{\csname bibitem#1\endcsname}%
\bibitem{littlewoodbes02}%
  \BibitemOpen
  \bibfield{author}{%
  \bibinfo {author} {\bibfnamefont{P.~B.}\ \bibnamefont{Littlewood}}, \bibinfo
  {author} {\bibfnamefont{G.~J.}\ \bibnamefont{Brown}}, \bibinfo {author}
  {\bibfnamefont{P.~R.}\ \bibnamefont{Eastham}},\ and\ \bibinfo {author}
  {\bibfnamefont{M.~H.}\ \bibnamefont{Szymanska}},\ }%
  \bibfield{journal}{%
  \bibinfo {journal} {Phys. Status Solidi B}\ }%
  \textbf{\bibinfo {volume} {234}},\ \bibinfo {pages} {36} (\bibinfo {year}
  {2002})%
  \bibAnnoteFile{NoStop}{littlewoodbes02}%
\bibitem{haugh67}%
  \BibitemOpen
  \bibfield{author}{%
  \bibinfo {author} {\bibfnamefont{H.}~\bibnamefont{Haug}}\ and\ \bibinfo
  {author} {\bibfnamefont{H.}~\bibnamefont{Haken}},\ }%
  \bibfield{journal}{%
  \bibinfo {journal} {Z. Phys. A: Hadrons Nucl.}\ }%
  \textbf{\bibinfo {volume} {204}},\ \bibinfo {pages} {262} (\bibinfo {year}
  {1967})%
  \bibAnnoteFile{NoStop}{haugh67}%
\bibitem{haugkoch}%
  \BibitemOpen
  \bibfield{author}{%
  \bibinfo {author} {\bibfnamefont{H.}~\bibnamefont{Haug}}\ and\ \bibinfo
  {author} {\bibfnamefont{S.~W.}\ \bibnamefont{Koch}},\ }%
  \emph{\bibinfo {title} {{Quantum theory of the optical and electronic
  properties of semiconductors}}},\ \bibinfo {edition} {5th}\ ed.\ (\bibinfo
  {publisher} {World Scientific},\ \bibinfo {address} {Singapore},\ \bibinfo
  {year} {2009})%
  \bibAnnoteFile{NoStop}{haugkoch}%
\bibitem{schmittrinkeh86}%
  \BibitemOpen
  \bibfield{author}{%
  \bibinfo {author} {\bibfnamefont{S.}~\bibnamefont{Schmitt-Rink}}, \bibinfo
  {author} {\bibfnamefont{C.}~\bibnamefont{Ell}},\ and\ \bibinfo {author}
  {\bibfnamefont{H.}~\bibnamefont{Haug}},\ }%
  \bibfield{journal}{%
  \Doi{10.1103/PhysRevB.33.1183}{\bibinfo {journal} {Phys. Rev. B}}\ }%
  \textbf{\bibinfo {volume} {33}},\ \bibinfo {pages} {1183} (\bibinfo {year}
  {1986})%
  \bibAnnoteFile{NoStop}{schmittrinkeh86}%
\bibitem{zhulittlewood96}%
  \BibitemOpen
  \bibfield{author}{%
  \bibinfo {author} {\bibfnamefont{X.}~\bibnamefont{Zhu}}, \bibinfo {author}
  {\bibfnamefont{M.~S.}\ \bibnamefont{Hybertsen}},\ and\ \bibinfo {author}
  {\bibfnamefont{P.~B.}\ \bibnamefont{Littlewood}},\ }%
  \bibfield{journal}{%
  \Doi{10.1103/PhysRevB.54.13575}{\bibinfo {journal} {Phys. Rev. B}}\ }%
  \textbf{\bibinfo {volume} {54}},\ \bibinfo {pages} {13575} (\bibinfo {year}
  {1996})%
  \bibAnnoteFile{NoStop}{zhulittlewood96}%
\bibitem{elliott57}%
  \BibitemOpen
  \bibfield{author}{%
  \bibinfo {author} {\bibfnamefont{R.~J.}\ \bibnamefont{Elliott}},\ }%
  \bibfield{journal}{%
  \bibinfo {journal} {Phys. Rev.}\ }%
  \textbf{\bibinfo {volume} {108}},\ \bibinfo {pages} {1384} (\bibinfo {year}
  {1957})%
  \bibAnnoteFile{NoStop}{elliott57}%
\bibitem{Ivanov}%
  \BibitemOpen
  \bibfield{author}{%
  \bibinfo {author} {\bibfnamefont{A.~L.}\ \bibnamefont{Ivanov}}, \bibinfo
  {author} {\bibfnamefont{H.}~\bibnamefont{Haug}},\ and\ \bibinfo {author}
  {\bibfnamefont{L.~V.}\ \bibnamefont{Keldysh}},\ }%
  \bibfield{journal}{%
  \bibinfo {journal} {Physics reports}\ }%
  \textbf{\bibinfo {volume} {296}},\ \bibinfo {pages} {237} (\bibinfo {year}
  {1998})%
  \bibAnnoteFile{NoStop}{Ivanov}%
\bibitem{shockleyr52}%
  \BibitemOpen
  \bibfield{author}{%
  \bibinfo {author} {\bibfnamefont{W.}~\bibnamefont{Shockley}}\ and\ \bibinfo
  {author} {\bibfnamefont{W.~T.}\ \bibnamefont{Read}},\ }%
  \bibfield{journal}{%
  \bibinfo {journal} {Phys. Rev.}\ }%
  \textbf{\bibinfo {volume} {87}},\ \bibinfo {pages} {835} (\bibinfo {year}
  {1952})%
  \bibAnnoteFile{NoStop}{shockleyr52}%
\bibitem{haug78}%
  \BibitemOpen
  \bibfield{author}{%
  \bibinfo {author} {\bibfnamefont{A.}~\bibnamefont{Haug}},\ }%
  \bibfield{journal}{%
  \bibinfo {journal} {Solid-State Electron.}\ }%
  \textbf{\bibinfo {volume} {21}},\ \bibinfo {pages} {1281} (\bibinfo {year}
  {1978})%
  \bibAnnoteFile{NoStop}{haug78}%
\bibitem{pincherle55}%
  \BibitemOpen
  \bibfield{author}{%
  \bibinfo {author} {\bibfnamefont{L.}~\bibnamefont{Pincherle}},\ }%
  \bibfield{journal}{%
  \bibinfo {journal} {Proc. Phys. Soc. London, Sect. B}\ }%
  \textbf{\bibinfo {volume} {68}},\ \bibinfo {pages} {319} (\bibinfo {year}
  {1955})%
  \bibAnnoteFile{NoStop}{pincherle55}%
\bibitem{beattiel59}%
  \BibitemOpen
  \bibfield{author}{%
  \bibinfo {author} {\bibfnamefont{A.~R.}\ \bibnamefont{Beattie}}\ and\
  \bibinfo {author} {\bibfnamefont{P.~T.}\ \bibnamefont{Landsberg}},\ }%
  \bibfield{journal}{%
  \bibinfo {journal} {Proc. R. Soc. London Ser. A}\ }%
  \textbf{\bibinfo {volume} {249}},\ \bibinfo {pages} {16} (\bibinfo {year}
  {1959})%
  \bibAnnoteFile{NoStop}{beattiel59}%
\bibitem{lius06}%
  \BibitemOpen
  \bibfield{author}{%
  \bibinfo {author} {\bibfnamefont{Y.}~\bibnamefont{Liu}}\ and\ \bibinfo
  {author} {\bibfnamefont{D.}~\bibnamefont{Snoke}},\ }%
  \bibfield{journal}{%
  \bibinfo {journal} {Solid State Commun.}\ }%
  \textbf{\bibinfo {volume} {140}},\ \bibinfo {pages} {208} (\bibinfo {year}
  {2006})%
  \bibAnnoteFile{NoStop}{lius06}%
\bibitem{kavoulakisbw96}%
  \BibitemOpen
  \bibfield{author}{%
  \bibinfo {author} {\bibfnamefont{G.~M.}\ \bibnamefont{Kavoulakis}}, \bibinfo
  {author} {\bibfnamefont{G.}~\bibnamefont{Baym}},\ and\ \bibinfo {author}
  {\bibfnamefont{J.~P.}\ \bibnamefont{Wolfe}},\ }%
  \bibfield{journal}{%
  \bibinfo {journal} {Phys. Rev. B}\ }%
  \textbf{\bibinfo {volume} {53}},\ \bibinfo {pages} {7227} (\bibinfo {year}
  {1996})%
  \bibAnnoteFile{NoStop}{kavoulakisbw96}%
\bibitem{kavoulakisb96}%
  \BibitemOpen
  \bibfield{author}{%
  \bibinfo {author} {\bibfnamefont{G.~M.}\ \bibnamefont{Kavoulakis}}\ and\
  \bibinfo {author} {\bibfnamefont{G.}~\bibnamefont{Baym}},\ }%
  \bibfield{journal}{%
  \bibinfo {journal} {Phys. Rev. B}\ }%
  \textbf{\bibinfo {volume} {54}},\ \bibinfo {pages} {16625} (\bibinfo {year}
  {1996})%
  \bibAnnoteFile{NoStop}{kavoulakisb96}%
\bibitem{jolkjk02}%
  \BibitemOpen
  \bibfield{author}{%
  \bibinfo {author} {\bibfnamefont{A.}~\bibnamefont{Jolk}}, \bibinfo {author}
  {\bibfnamefont{M.}~\bibnamefont{J{\"o}rger}},\ and\ \bibinfo {author}
  {\bibfnamefont{C.}~\bibnamefont{Klingshirn}},\ }%
  \bibfield{journal}{%
  \bibinfo {journal} {Phys. Rev. B}\ }%
  \textbf{\bibinfo {volume} {65}},\ \bibinfo {pages} {245209} (\bibinfo {year}
  {2002})%
  \bibAnnoteFile{NoStop}{jolkjk02}%
\bibitem{bassani75}%
  \BibitemOpen
  \bibfield{author}{%
  \bibinfo {author} {\bibfnamefont{F.}~\bibnamefont{Bassani}}\ and\ \bibinfo
  {author} {\bibfnamefont{G.}~\bibnamefont{Pastori~Parravicini}},\ }%
  \emph{\bibinfo {title} {Electronic states and optical transitions in
  solids}}\ (\bibinfo {publisher} {Pergamon},\ \bibinfo {year} {1975})%
  \bibAnnoteFile{NoStop}{bassani75}%
\bibitem{andreani95}%
  \BibitemOpen
  \bibfield{author}{%
  \bibinfo {author} {\bibfnamefont{L.~C.}\ \bibnamefont{Andreani}},\ }%
  in\ \emph{\bibinfo {booktitle} {Confined electrons and photons: new physics
  and applications}},\ \bibinfo {editor} {edited by\ \bibinfo {editor}
  {\bibfnamefont{E.}~\bibnamefont{Burstein}}\ and\ \bibinfo {editor}
  {\bibfnamefont{C.}~\bibnamefont{Weisbuch}}}\ (\bibinfo {publisher} {Plenum
  Publishing Corporation},\ \bibinfo {year} {1995})\ p.~\bibinfo {pages} {57}%
  \bibAnnoteFile{NoStop}{andreani95}%
\bibitem{andreanibassani}%
  \BibitemOpen
  \bibfield{author}{%
  \bibinfo {author} {\bibfnamefont{L.~C.}\ \bibnamefont{Andreani}}\ and\
  \bibinfo {author} {\bibfnamefont{F.}~\bibnamefont{Bassani}},\ }%
  \bibfield{journal}{%
  \bibinfo {journal} {Phys. Rev. B}\ }%
  \textbf{\bibinfo {volume} {41}},\ \bibinfo {pages} {7536} (\bibinfo {year}
  {1990})%
  \bibAnnoteFile{NoStop}{andreanibassani}%
\bibitem{drummond_jastrow_2004}%
  \BibitemOpen
  \bibfield{author}{%
  \bibinfo {author} {\bibfnamefont{N.~D.}\ \bibnamefont{Drummond}}, \bibinfo
  {author} {\bibfnamefont{M.~D.}\ \bibnamefont{Towler}},\ and\ \bibinfo
  {author} {\bibfnamefont{R.~J.}\ \bibnamefont{Needs}},\ }%
  \bibfield{journal}{%
  \bibinfo {journal} {Phys. Rev. B}\ }%
  \textbf{\bibinfo {volume} {70}},\ \bibinfo {pages} {235119} (\bibinfo {year}
  {2004})%
  \bibAnnoteFile{NoStop}{drummond_jastrow_2004}%
\bibitem{kato_cusp_1957}%
  \BibitemOpen
  \bibfield{author}{%
  \bibinfo {author} {\bibfnamefont{T.}~\bibnamefont{Kato}},\ }%
  \bibfield{journal}{%
  \bibinfo {journal} {Commun. Pure Appl. Math.}\ }%
  \textbf{\bibinfo {volume} {10}},\ \bibinfo {pages} {151} (\bibinfo {year}
  {1957})%
  \bibAnnoteFile{NoStop}{kato_cusp_1957}%
\bibitem{umrigar_emin_2007}%
  \BibitemOpen
  \bibfield{author}{%
  \bibinfo {author} {\bibfnamefont{J.}~\bibnamefont{Toulouse}}\ and\ \bibinfo
  {author} {\bibfnamefont{C.~J.}\ \bibnamefont{Umrigar}},\ }%
  \bibfield{journal}{%
  \bibinfo {journal} {J. Chem. Phys.}\ }%
  \textbf{\bibinfo {volume} {126}},\ \bibinfo {pages} {084102} (\bibinfo {year}
  {2007})%
  \bibAnnoteFile{NoStop}{umrigar_emin_2007}%
\bibitem{needs_casino_2010}%
  \BibitemOpen
  \bibfield{author}{%
  \bibinfo {author} {\bibfnamefont{R.~J.}\ \bibnamefont{Needs}}, \bibinfo
  {author} {\bibfnamefont{M.~D.}\ \bibnamefont{Towler}}, \bibinfo {author}
  {\bibfnamefont{N.~D.}\ \bibnamefont{Drummond}},\ and\ \bibinfo {author}
  {\bibfnamefont{P.}~\bibnamefont{L\'opez R\'\i~os}},\ }%
  \bibfield{journal}{%
  \bibinfo {journal} {J. Phys.: Condens. Matter}\ }%
  \textbf{\bibinfo {volume} {22}},\ \bibinfo {pages} {023201} (\bibinfo {year}
  {2010})%
  \bibAnnoteFile{NoStop}{needs_casino_2010}%
\bibitem{leedn09}%
  \BibitemOpen
  \bibfield{author}{%
  \bibinfo {author} {\bibfnamefont{R.~M.}\ \bibnamefont{Lee}}, \bibinfo
  {author} {\bibfnamefont{N.~D.}\ \bibnamefont{Drummond}},\ and\ \bibinfo
  {author} {\bibfnamefont{R.~J.}\ \bibnamefont{Needs}},\ }%
  \bibfield{journal}{%
  \bibinfo {journal} {Phys. Rev. B}\ }%
  \textbf{\bibinfo {volume} {79}},\ \bibinfo {pages} {125308} (\bibinfo {year}
  {2009})%
  \bibAnnoteFile{NoStop}{leedn09}%
\bibitem{Chuang}%
  \BibitemOpen
  \bibfield{author}{%
  \bibinfo {author} {\bibfnamefont{S.~L.}\ \bibnamefont{Chuang}},\ }%
  \emph{\bibinfo {title} {{Physics of optoelectronic devices}}}\ (\bibinfo
  {publisher} {Wiley New York},\ \bibinfo {year} {1995})%
  \bibAnnoteFile{NoStop}{Chuang}%
\bibitem{Schiff}%
  \BibitemOpen
  \bibfield{author}{%
  \bibinfo {author} {\bibfnamefont{L.~I.}\ \bibnamefont{Schiff}},\ }%
  \emph{\bibinfo {title} {Quantum Mechanics}}\ (\bibinfo {publisher} {Mc
  Graw-Hill Inc.},\ \bibinfo {year} {1968})%
  \bibAnnoteFile{NoStop}{Schiff}%
\bibitem{akiyamakmk90}%
  \BibitemOpen
  \bibfield{author}{%
  \bibinfo {author} {\bibfnamefont{H.}~\bibnamefont{Akiyama}}, \bibinfo
  {author} {\bibfnamefont{T.}~\bibnamefont{Kuga}}, \bibinfo {author}
  {\bibfnamefont{M.}~\bibnamefont{Matsuoka}},\ and\ \bibinfo {author}
  {\bibfnamefont{M.}~\bibnamefont{Kuwata-Gonokami}},\ }%
  \bibfield{journal}{%
  \Doi{10.1103/PhysRevB.42.5621}{\bibinfo {journal} {Phys. Rev. B}}\ }%
  \textbf{\bibinfo {volume} {42}},\ \bibinfo {pages} {5621} (\bibinfo {year}
  {1990})%
  \bibAnnoteFile{NoStop}{akiyamakmk90}%
\bibitem{nagaikg02}%
  \BibitemOpen
  \bibfield{author}{%
  \bibinfo {author} {\bibfnamefont{M.}~\bibnamefont{Nagai}}\ and\ \bibinfo
  {author} {\bibfnamefont{M.}~\bibnamefont{Kuwata-Gonokami}},\ }%
  \bibfield{journal}{%
  \bibinfo {journal} {J. Lumin.}\ }%
  \textbf{\bibinfo {volume} {100}},\ \bibinfo {pages} {233} (\bibinfo {year}
  {2002})%
  \bibAnnoteFile{NoStop}{nagaikg02}%
\bibitem{herzp02}%
  \BibitemOpen
  \bibfield{author}{%
  \bibinfo {author} {\bibfnamefont{L.~M.}\ \bibnamefont{Herz}}\ and\ \bibinfo
  {author} {\bibfnamefont{R.~T.}\ \bibnamefont{Phillips}},\ }%
  \bibfield{journal}{%
  \bibinfo {journal} {Nat. Mater.}\ }%
  \textbf{\bibinfo {volume} {1}},\ \bibinfo {pages} {212} (\bibinfo {year}
  {2002})%
  \bibAnnoteFile{NoStop}{herzp02}%
\bibitem{Note1}%
  \BibitemOpen
  \bibinfo {note} {Equation (\ref {eq:polariton}) calculates the general decay
  rate into a two-polariton state and not the rate at which photons are emitted
  from the semiconductor, since it is this general decay rate that is
  experimentally accessible.}%
  \bibAnnoteFile{Stop}{Note1}%
\bibitem{jfprivate}%
  \BibitemOpen
  \bibfield{author}{%
  \bibinfo {author} {\bibfnamefont{J.~J.}\ \bibnamefont{Finley}},\ }%
  \bibinfo {howpublished} {Private Communication} (\bibinfo {year} {2012})%
  \bibAnnoteFile{NoStop}{jfprivate}%
\bibitem{phillipslovering}%
  \BibitemOpen
  \bibfield{author}{%
  \bibinfo {author} {\bibfnamefont{R.~T.}\ \bibnamefont{Phillips}}, \bibinfo
  {author} {\bibfnamefont{D.~J.}\ \bibnamefont{Lovering}}, \bibinfo {author}
  {\bibfnamefont{G.~J.}\ \bibnamefont{Denton}},\ and\ \bibinfo {author}
  {\bibfnamefont{G.~W.}\ \bibnamefont{Smith}},\ }%
  \bibfield{journal}{%
  \bibinfo {journal} {Phys. Rev. B}\ }%
  \textbf{\bibinfo {volume} {45}},\ \bibinfo {pages} {4308} (\bibinfo {year}
  {1992})%
  \bibAnnoteFile{NoStop}{phillipslovering}%
\bibitem{spiegel}%
  \BibitemOpen
  \bibfield{author}{%
  \bibinfo {author} {\bibfnamefont{R.}~\bibnamefont{Spiegel}}, \bibinfo
  {author} {\bibfnamefont{G.}~\bibnamefont{Bacher}}, \bibinfo {author}
  {\bibfnamefont{A.}~\bibnamefont{Forchel}}, \bibinfo {author}
  {\bibfnamefont{B.}~\bibnamefont{Jobst}}, \bibinfo {author}
  {\bibfnamefont{D.}~\bibnamefont{Hommel}},\ and\ \bibinfo {author}
  {\bibfnamefont{G.}~\bibnamefont{Landwehr}},\ }%
  \bibfield{journal}{%
  \bibinfo {journal} {Phys. Rev. B}\ }%
  \textbf{\bibinfo {volume} {55}},\ \bibinfo {pages} {9866} (\bibinfo {year}
  {1997})%
  \bibAnnoteFile{NoStop}{spiegel}%
\bibitem{charbonneau}%
  \BibitemOpen
  \bibfield{author}{%
  \bibinfo {author} {\bibfnamefont{S.}~\bibnamefont{Charbonneau}}, \bibinfo
  {author} {\bibfnamefont{T.}~\bibnamefont{Steiner}}, \bibinfo {author}
  {\bibfnamefont{M.~L.~W.}\ \bibnamefont{Thewalt}}, \bibinfo {author}
  {\bibfnamefont{E.~S.}\ \bibnamefont{Koteles}}, \bibinfo {author}
  {\bibfnamefont{J.~Y.}\ \bibnamefont{Chi}},\ and\ \bibinfo {author}
  {\bibfnamefont{B.}~\bibnamefont{Elman}},\ }%
  \bibfield{journal}{%
  \bibinfo {journal} {Phys. Rev. B}\ }%
  \textbf{\bibinfo {volume} {38}},\ \bibinfo {pages} {3583} (\bibinfo {year}
  {1988})%
  \bibAnnoteFile{NoStop}{charbonneau}%
\bibitem{devaud}%
  \BibitemOpen
  \bibfield{author}{%
  \bibinfo {author} {\bibfnamefont{B.}~\bibnamefont{Deveaud}}, \bibinfo
  {author} {\bibfnamefont{F.}~\bibnamefont{Cl\'erot}}, \bibinfo {author}
  {\bibfnamefont{N.}~\bibnamefont{Roy}}, \bibinfo {author}
  {\bibfnamefont{K.}~\bibnamefont{Satzke}}, \bibinfo {author}
  {\bibfnamefont{B.}~\bibnamefont{Sermage}},\ and\ \bibinfo {author}
  {\bibfnamefont{D.~S.}\ \bibnamefont{Katzer}},\ }%
  \bibfield{journal}{%
  \bibinfo {journal} {Phys. Rev. Lett.}\ }%
  \textbf{\bibinfo {volume} {67}},\ \bibinfo {pages} {2355} (\bibinfo {year}
  {1991})%
  \bibAnnoteFile{NoStop}{devaud}%
\bibitem{cingolani}%
  \BibitemOpen
  \bibfield{author}{%
  \bibinfo {author} {\bibfnamefont{R.}~\bibnamefont{Cingolani}}, \bibinfo
  {author} {\bibfnamefont{K.}~\bibnamefont{Ploog}}, \bibinfo {author}
  {\bibfnamefont{G.}~\bibnamefont{Peter}}, \bibinfo {author}
  {\bibfnamefont{R.}~\bibnamefont{Hahn}}, \bibinfo {author}
  {\bibfnamefont{E.~O.}\ \bibnamefont{G\"obel}}, \bibinfo {author}
  {\bibfnamefont{C.}~\bibnamefont{Moro}},\ and\ \bibinfo {author}
  {\bibfnamefont{A.}~\bibnamefont{Cingolani}},\ }%
  \bibfield{journal}{%
  \bibinfo {journal} {Phys. Rev. B}\ }%
  \textbf{\bibinfo {volume} {41}},\ \bibinfo {pages} {3272} (\bibinfo {year}
  {1990})%
  \bibAnnoteFile{NoStop}{cingolani}%
\bibitem{steiner89}%
  \BibitemOpen
  \bibfield{author}{%
  \bibinfo {author} {\bibfnamefont{T.~W.}\ \bibnamefont{Steiner}}, \bibinfo
  {author} {\bibfnamefont{A.~G.}\ \bibnamefont{Steele}}, \bibinfo {author}
  {\bibfnamefont{S.}~\bibnamefont{Charbonneau}}, \bibinfo {author}
  {\bibfnamefont{M.~L.~W.}\ \bibnamefont{Thewalt}}, \bibinfo {author}
  {\bibfnamefont{E.~S.}\ \bibnamefont{Koteles}},\ and\ \bibinfo {author}
  {\bibfnamefont{B.}~\bibnamefont{Elman}},\ }%
  \bibfield{journal}{%
  \bibinfo {journal} {Solid State Commun.}\ }%
  \textbf{\bibinfo {volume} {69}},\ \bibinfo {pages} {1139} (\bibinfo {year}
  {1989})%
  \bibAnnoteFile{NoStop}{steiner89}%
\bibitem{yamada95lifetime}%
  \BibitemOpen
  \bibfield{author}{%
  \bibinfo {author} {\bibfnamefont{Y.}~\bibnamefont{Yamada}}, \bibinfo {author}
  {\bibfnamefont{T.}~\bibnamefont{Mishina}}, \bibinfo {author}
  {\bibfnamefont{Y.}~\bibnamefont{Masumoto}}, \bibinfo {author}
  {\bibfnamefont{Y.}~\bibnamefont{Kawakami}}, \bibinfo {author}
  {\bibfnamefont{S.}~\bibnamefont{Yamaguchi}}, \bibinfo {author}
  {\bibfnamefont{K.}~\bibnamefont{Ichino}}, \bibinfo {author}
  {\bibfnamefont{S.}~\bibnamefont{Fujita}}, \bibinfo {author}
  {\bibfnamefont{S.}~\bibnamefont{Fujita}},\ and\ \bibinfo {author}
  {\bibfnamefont{T.}~\bibnamefont{Taguchi}},\ }%
  \bibfield{journal}{%
  \bibinfo {journal} {Phys. Rev. B}\ }%
  \textbf{\bibinfo {volume} {51}},\ \bibinfo {pages} {2596} (\bibinfo {year}
  {1995})%
  \bibAnnoteFile{NoStop}{yamada95lifetime}%
\bibitem{hasuoCuCl}%
  \BibitemOpen
  \bibfield{author}{%
  \bibinfo {author} {\bibfnamefont{M.}~\bibnamefont{Hasuo}}, \bibinfo {author}
  {\bibfnamefont{M.}~\bibnamefont{Nishino}},\ and\ \bibinfo {author}
  {\bibfnamefont{N.}~\bibnamefont{Nagasawa}},\ }%
  \bibfield{journal}{%
  \bibinfo {journal} {J. Luminescence}\ }%
  \textbf{\bibinfo {volume} {60}},\ \bibinfo {pages} {672} (\bibinfo {year}
  {1994})%
  \bibAnnoteFile{NoStop}{hasuoCuCl}%
\bibitem{IvanovCuClexp}%
  \BibitemOpen
  \bibfield{author}{%
  \bibinfo {author} {\bibfnamefont{A.~L.}\ \bibnamefont{Ivanov}}, \bibinfo
  {author} {\bibfnamefont{M.}~\bibnamefont{Hasuo}}, \bibinfo {author}
  {\bibfnamefont{N.}~\bibnamefont{Nagasawa}},\ and\ \bibinfo {author}
  {\bibfnamefont{H.}~\bibnamefont{Haug}},\ }%
  \bibfield{journal}{%
  \bibinfo {journal} {Phys. Rev. B}\ }%
  \textbf{\bibinfo {volume} {52}},\ \bibinfo {pages} {11017} (\bibinfo {year}
  {1995})%
  \bibAnnoteFile{NoStop}{IvanovCuClexp}%
\bibitem{edamatsu96}%
  \BibitemOpen
  \bibfield{author}{%
  \bibinfo {author} {\bibfnamefont{K.}~\bibnamefont{Edamatsu}},\ }%
  \bibfield{journal}{%
  \bibinfo {journal} {J. Luminescence}\ }%
  \textbf{\bibinfo {volume} {70}},\ \bibinfo {pages} {377} (\bibinfo {year}
  {1996})%
  \bibAnnoteFile{NoStop}{edamatsu96}%
\bibitem{ulrichf05}%
  \BibitemOpen
  \bibfield{author}{%
  \bibinfo {author} {\bibfnamefont{S.~M.}\ \bibnamefont{Ulrich}}, \bibinfo
  {author} {\bibfnamefont{M.}~\bibnamefont{Benyoucef}}, \bibinfo {author}
  {\bibfnamefont{P.}~\bibnamefont{Michler}}, \bibinfo {author}
  {\bibfnamefont{N.}~\bibnamefont{Baer}}, \bibinfo {author}
  {\bibfnamefont{P.}~\bibnamefont{Gartner}}, \bibinfo {author}
  {\bibfnamefont{F.}~\bibnamefont{Jahnke}}, \bibinfo {author}
  {\bibfnamefont{M.}~\bibnamefont{Schwab}}, \bibinfo {author}
  {\bibfnamefont{H.}~\bibnamefont{Kurtze}}, \bibinfo {author}
  {\bibfnamefont{M.}~\bibnamefont{Bayer}}, \bibinfo {author}
  {\bibfnamefont{S.}~\bibnamefont{Fafard}}, \bibinfo {author}
  {\bibfnamefont{Z.}~\bibnamefont{Wasilewski}},\ and\ \bibinfo {author}
  {\bibfnamefont{A.}~\bibnamefont{Forchel}},\ }%
  \bibfield{journal}{%
  \Doi{10.1103/PhysRevB.71.235328}{\bibinfo {journal} {Phys. Rev. B}}\ }%
  \textbf{\bibinfo {volume} {71}},\ \bibinfo {pages} {235328} (\bibinfo {year}
  {2005})%
  \bibAnnoteFile{NoStop}{ulrichf05}%
\bibitem{narvaezGBZ05}%
  \BibitemOpen
  \bibfield{author}{%
  \bibinfo {author} {\bibfnamefont{G.~A.}\ \bibnamefont{Narvaez}}, \bibinfo
  {author} {\bibfnamefont{G.}~\bibnamefont{Bester}},\ and\ \bibinfo {author}
  {\bibfnamefont{A.}~\bibnamefont{Zunger}},\ }%
  \bibfield{journal}{%
  \Doi{10.1103/PhysRevB.72.245318}{\bibinfo {journal} {Phys. Rev. B}}\ }%
  \textbf{\bibinfo {volume} {72}},\ \bibinfo {pages} {245318} (\bibinfo {year}
  {2005})%
  \bibAnnoteFile{NoStop}{narvaezGBZ05}%
\bibitem{wimmer06}%
  \BibitemOpen
  \bibfield{author}{%
  \bibinfo {author} {\bibfnamefont{M.}~\bibnamefont{Wimmer}}, \bibinfo {author}
  {\bibfnamefont{S.~V.}\ \bibnamefont{Nair}},\ and\ \bibinfo {author}
  {\bibfnamefont{J.}~\bibnamefont{Shumway}},\ }%
  \bibfield{journal}{%
  \bibinfo {journal} {Phys. Rev. B}\ }%
  \textbf{\bibinfo {volume} {73}},\ \bibinfo {pages} {165305} (\bibinfo {year}
  {2006})%
  \bibAnnoteFile{NoStop}{wimmer06}%
\bibitem{kuther98}%
  \BibitemOpen
  \bibfield{author}{%
  \bibinfo {author} {\bibfnamefont{A.}~\bibnamefont{Kuther}}, \bibinfo {author}
  {\bibfnamefont{M.}~\bibnamefont{Bayer}}, \bibinfo {author}
  {\bibfnamefont{A.}~\bibnamefont{Forchel}}, \bibinfo {author}
  {\bibfnamefont{A.}~\bibnamefont{Gorbunov}}, \bibinfo {author}
  {\bibfnamefont{V.~B.}\ \bibnamefont{Timofeev}}, \bibinfo {author}
  {\bibfnamefont{F.}~\bibnamefont{Sch{\"a}fer}},\ and\ \bibinfo {author}
  {\bibfnamefont{J.~P.}\ \bibnamefont{Reithmaier}},\ }%
  \bibfield{journal}{%
  \bibinfo {journal} {Phys. Rev. B}\ }%
  \textbf{\bibinfo {volume} {58}},\ \bibinfo {pages} {R7508} (\bibinfo {year}
  {1998})%
  \bibAnnoteFile{NoStop}{kuther98}%
\bibitem{kuroda06}%
  \BibitemOpen
  \bibfield{author}{%
  \bibinfo {author} {\bibfnamefont{K.}~\bibnamefont{Kuroda}}, \bibinfo {author}
  {\bibfnamefont{T.}~\bibnamefont{Kuroda}}, \bibinfo {author}
  {\bibfnamefont{K.}~\bibnamefont{Sakoda}}, \bibinfo {author}
  {\bibfnamefont{K.}~\bibnamefont{Watanabe}}, \bibinfo {author}
  {\bibfnamefont{N.}~\bibnamefont{Koguchi}},\ and\ \bibinfo {author}
  {\bibfnamefont{G.}~\bibnamefont{Kido}},\ }%
  \bibfield{journal}{%
  \bibinfo {journal} {Appl. Phys. Lett.}\ }%
  \textbf{\bibinfo {volume} {88}},\ \bibinfo {pages} {124101} (\bibinfo {year}
  {2006})%
  \bibAnnoteFile{NoStop}{kuroda06}%
\bibitem{yamada95binding}%
  \BibitemOpen
  \bibfield{author}{%
  \bibinfo {author} {\bibfnamefont{Y.}~\bibnamefont{Yamada}}, \bibinfo {author}
  {\bibfnamefont{T.}~\bibnamefont{Mishina}}, \bibinfo {author}
  {\bibfnamefont{Y.}~\bibnamefont{Masumoto}}, \bibinfo {author}
  {\bibfnamefont{Y.}~\bibnamefont{Kawakami}}, \bibinfo {author}
  {\bibfnamefont{J.}~\bibnamefont{Suda}}, \bibinfo {author}
  {\bibfnamefont{S.}~\bibnamefont{Fujita}},\ and\ \bibinfo {author}
  {\bibfnamefont{S.}~\bibnamefont{Fujita}},\ }%
  \bibfield{journal}{%
  \bibinfo {journal} {Phys. Rev. B}\ }%
  \textbf{\bibinfo {volume} {52}},\ \bibinfo {pages} {R2289} (\bibinfo {year}
  {1995})%
  \bibAnnoteFile{NoStop}{yamada95binding}%
\bibitem{Ueta}%
  \BibitemOpen
  \bibfield{author}{%
  \bibinfo {author} {\bibfnamefont{M.}~\bibnamefont{Ueta}}, \bibinfo {author}
  {\bibfnamefont{H.}~\bibnamefont{Kanzaki}}, \bibinfo {author}
  {\bibfnamefont{K.}~\bibnamefont{Kobayashi}}, \bibinfo {author}
  {\bibfnamefont{Y.}~\bibnamefont{Toyozawa}},\ and\ \bibinfo {author}
  {\bibfnamefont{E.}~\bibnamefont{Hanamura}},\ }%
  \emph{\bibinfo {title} {Excitonic Processes in Solids}},\ Vol.~\bibinfo
  {volume} {60}\ (\bibinfo {publisher} {Springer Series in Solid-State
  Sciences},\ \bibinfo {year} {1986})%
  \bibAnnoteFile{NoStop}{Ueta}%
\bibitem{itoh1989nonlinear}%
  \BibitemOpen
  \bibfield{author}{%
  \bibinfo {author} {\bibfnamefont{T.}~\bibnamefont{Itoh}}, \bibinfo {author}
  {\bibfnamefont{F.}~\bibnamefont{Jin}}, \bibinfo {author}
  {\bibfnamefont{Y.}~\bibnamefont{Iwabuchi}},\ and\ \bibinfo {author}
  {\bibfnamefont{T.}~\bibnamefont{Ikehara}},\ }%
  \bibfield{journal}{%
  \bibinfo {journal} {Springer Proceedings in Physics}\ }%
  \textbf{\bibinfo {volume} {36}},\ \bibinfo {pages} {76} (\bibinfo {year}
  {1989})%
  \bibAnnoteFile{NoStop}{itoh1989nonlinear}%
\bibitem{Park00}%
  \BibitemOpen
  \bibfield{author}{%
  \bibinfo {author} {\bibfnamefont{S.}~\bibnamefont{Park}}, \bibinfo {author}
  {\bibfnamefont{G.}~\bibnamefont{Jeen}}, \bibinfo {author}
  {\bibfnamefont{H.}~\bibnamefont{Kim}},\ and\ \bibinfo {author}
  {\bibfnamefont{I.}~\bibnamefont{Kim}},\ }%
  \bibfield{journal}{%
  \bibinfo {journal} {Journal of the Korean Physical Society}\ }%
  \textbf{\bibinfo {volume} {37}},\ \bibinfo {pages} {309} (\bibinfo {year}
  {2000})%
  \bibAnnoteFile{NoStop}{Park00}%
\bibitem{nozue}%
  \BibitemOpen
  \bibfield{author}{%
  \bibinfo {author} {\bibfnamefont{Y.}~\bibnamefont{Nozue}}, \bibinfo {author}
  {\bibfnamefont{M.}~\bibnamefont{Itoh}},\ and\ \bibinfo {author}
  {\bibfnamefont{K.}~\bibnamefont{Cho}},\ }%
  \bibfield{journal}{%
  \bibinfo {journal} {Journal of the Physical Society of Japan}\ }%
  \textbf{\bibinfo {volume} {50}},\ \bibinfo {pages} {889} (\bibinfo {year}
  {1981})%
  \bibAnnoteFile{NoStop}{nozue}%
\bibitem{AndreanidelSole}%
  \BibitemOpen
  \bibfield{author}{%
  \bibinfo {author} {\bibfnamefont{L.~C.}\ \bibnamefont{Andreani}}, \bibinfo
  {author} {\bibfnamefont{A.}~\bibnamefont{D'Andrea}},\ and\ \bibinfo {author}
  {\bibfnamefont{R.}~\bibnamefont{Del~Sole}},\ }%
  \bibfield{journal}{%
  \bibinfo {journal} {Phys. Lett. A}\ }%
  \textbf{\bibinfo {volume} {168}},\ \bibinfo {pages} {451} (\bibinfo {year}
  {1992})%
  \bibAnnoteFile{NoStop}{AndreanidelSole}%
\bibitem{Singh}%
  \BibitemOpen
  \bibfield{author}{%
  \bibinfo {author} {\bibfnamefont{J.}~\bibnamefont{Singh}},\ }%
  \emph{\bibinfo {title} {Electronic and optoelectronic properties of
  semiconductor structures}}\ (\bibinfo {publisher} {Cambridge University
  Press},\ \bibinfo {year} {2003})%
  \bibAnnoteFile{NoStop}{Singh}%
\bibitem{stiergm99}%
  \BibitemOpen
  \bibfield{author}{%
  \bibinfo {author} {\bibfnamefont{O.}~\bibnamefont{Stier}}, \bibinfo {author}
  {\bibfnamefont{M.}~\bibnamefont{Grundmann}},\ and\ \bibinfo {author}
  {\bibfnamefont{D.}~\bibnamefont{Bimberg}},\ }%
  \bibfield{journal}{%
  \Doi{10.1103/PhysRevB.59.5688}{\bibinfo {journal} {Phys. Rev. B}}\ }%
  \textbf{\bibinfo {volume} {59}},\ \bibinfo {pages} {5688} (\bibinfo {year}
  {1999})%
  \bibAnnoteFile{NoStop}{stiergm99}%
\bibitem{Adachi}%
  \BibitemOpen
  \bibfield{author}{%
  \bibinfo {author} {\bibfnamefont{S.}~\bibnamefont{Adachi}},\ }%
  \emph{\bibinfo {title} {Physical Properties of III-V Semiconductor
  Compounds}}\ (\bibinfo {publisher} {Wiley Online Library},\ \bibinfo {year}
  {1992})%
  \bibAnnoteFile{NoStop}{Adachi}%
\bibitem{liao11}%
  \BibitemOpen
  \bibfield{author}{%
  \bibinfo {author} {\bibfnamefont{Y.-H.}\ \bibnamefont{Liao}}, \bibinfo
  {author} {\bibfnamefont{J.~I.}\ \bibnamefont{Climente}},\ and\ \bibinfo
  {author} {\bibfnamefont{S.-J.}\ \bibnamefont{Cheng}},\ }%
  \bibfield{journal}{%
  \bibinfo {journal} {Phys. Rev. B}\ }%
  \textbf{\bibinfo {volume} {83}},\ \bibinfo {pages} {165317} (\bibinfo {year}
  {2011})%
  \bibAnnoteFile{NoStop}{liao11}%
\bibitem{laucht}%
  \BibitemOpen
  \bibfield{author}{%
  \bibinfo {author} {\bibfnamefont{A.}~\bibnamefont{Laucht}}, \bibinfo {author}
  {\bibfnamefont{N.}~\bibnamefont{Hauke}}, \bibinfo {author}
  {\bibfnamefont{J.~M.}~\bibnamefont{Villas-B{\^o}as}}, \bibinfo {author}
  {\bibfnamefont{F.}~\bibnamefont{Hofbauer}}, \bibinfo {author}
  {\bibfnamefont{G.}~\bibnamefont{B{\"o}hm}}, \bibinfo {author}
  {\bibfnamefont{M.}~\bibnamefont{Kaniber}},\ and\ \bibinfo {author}
  {\bibfnamefont{J.~J.}\ \bibnamefont{Finley}},\ }%
  \bibfield{journal}{%
  \bibinfo {journal} {Phys. Rev. Lett.}\ }%
  \textbf{\bibinfo {volume} {103}},\ \bibinfo {pages} {087405} (\bibinfo {year}
  {2009})%
  \bibAnnoteFile{NoStop}{laucht}%
\bibitem{willatzen}%
  \BibitemOpen
  \bibfield{author}{%
  \bibinfo {author} {\bibfnamefont{M.}~\bibnamefont{Willatzen}}, \bibinfo
  {author} {\bibfnamefont{M.}~\bibnamefont{Cardona}},\ and\ \bibinfo {author}
  {\bibfnamefont{N.~E.}\ \bibnamefont{Christensen}},\ }%
  \bibfield{journal}{%
  \bibinfo {journal} {Phys. Rev. B}\ }%
  \textbf{\bibinfo {volume} {51}},\ \bibinfo {pages} {17992} (\bibinfo {year}
  {1995})%
  \bibAnnoteFile{NoStop}{willatzen}%
\bibitem{NairBeCuCl}%
  \BibitemOpen
  \bibfield{author}{%
  \bibinfo {author} {\bibfnamefont{S.~V.}\ \bibnamefont{Nair}}\ and\ \bibinfo
  {author} {\bibfnamefont{T.}~\bibnamefont{Takagahara}},\ }%
  \bibfield{journal}{%
  \bibinfo {journal} {Phys. Rev. B}\ }%
  \textbf{\bibinfo {volume} {55}},\ \bibinfo {pages} {5153} (\bibinfo {year}
  {1997})%
  \bibAnnoteFile{NoStop}{NairBeCuCl}%
\bibitem{masumotoBeCuCl}%
  \BibitemOpen
  \bibfield{author}{%
  \bibinfo {author} {\bibfnamefont{Y.}~\bibnamefont{Masumoto}}, \bibinfo
  {author} {\bibfnamefont{S.}~\bibnamefont{Okamoto}},\ and\ \bibinfo {author}
  {\bibfnamefont{S.}~\bibnamefont{Katayanagi}},\ }%
  \bibfield{journal}{%
  \bibinfo {journal} {Phys. Rev. B}\ }%
  \textbf{\bibinfo {volume} {50}},\ \bibinfo {pages} {18658} (\bibinfo {year}
  {1994})%
  \bibAnnoteFile{NoStop}{masumotoBeCuCl}%
\bibitem{bulaevloss05}%
  \BibitemOpen
  \bibfield{author}{%
  \bibinfo {author} {\bibfnamefont{D.~V.}\ \bibnamefont{Bulaev}}\ and\ \bibinfo
  {author} {\bibfnamefont{D.}~\bibnamefont{Loss}},\ }%
  \bibfield{journal}{%
  \bibinfo {journal} {Phys. Rev. Lett.}\ }%
  \textbf{\bibinfo {volume} {95}},\ \bibinfo {pages} {076805} (\bibinfo {year}
  {2005})%
  \bibAnnoteFile{NoStop}{bulaevloss05}%
\bibitem{chan86}%
  \BibitemOpen
  \bibfield{author}{%
  \bibinfo {author} {\bibfnamefont{K.~S.}\ \bibnamefont{Chan}},\ }%
  \bibfield{journal}{%
  \bibinfo {journal} {Journal of Physics C: Solid State Physics}\ }%
  \textbf{\bibinfo {volume} {19}},\ \bibinfo {pages} {L125} (\bibinfo {year}
  {1986})%
  \bibAnnoteFile{NoStop}{chan86}%
\bibitem{wimbauer94}%
  \BibitemOpen
  \bibfield{author}{%
  \bibinfo {author} {\bibfnamefont{T.}~\bibnamefont{Wimbauer}}, \bibinfo
  {author} {\bibfnamefont{K.}~\bibnamefont{Oettinger}}, \bibinfo {author}
  {\bibfnamefont{A.~L.}\ \bibnamefont{Efros}}, \bibinfo {author}
  {\bibfnamefont{B.~K.}\ \bibnamefont{Meyer}},\ and\ \bibinfo {author}
  {\bibfnamefont{H.}~\bibnamefont{Brugger}},\ }%
  \bibfield{journal}{%
  \bibinfo {journal} {Phys. Rev. B}\ }%
  \textbf{\bibinfo {volume} {50}},\ \bibinfo {pages} {8889} (\bibinfo {year}
  {1994})%
  \bibAnnoteFile{NoStop}{wimbauer94}%
\bibitem{hanamura}%
  \BibitemOpen
  \bibfield{author}{%
  \bibinfo {author} {\bibfnamefont{E.}~\bibnamefont{Hanamura}},\ }%
  \bibfield{journal}{%
  \bibinfo {journal} {Solid State Commun.}\ }%
  \textbf{\bibinfo {volume} {12}},\ \bibinfo {pages} {951} (\bibinfo {year}
  {1973})%
  \bibAnnoteFile{NoStop}{hanamura}%
\bibitem{Note2}%
  \BibitemOpen
  \bibinfo {note} {We note that the lengthscales $a_B$ and $a_E$ are only
  appropriate for describing this problem if the biexciton can be envisaged as
  being made up of two excitons that retain their identity in the bound state.
  This picture of a biexciton has been found to be correct at least in two
  dimensional systems with large electron-hole layer separation \cite
  {leedn09,szimmermann08, mfogler08}.}%
  \bibAnnoteFile{Stop}{Note2}%
\bibitem{banyaigeh87}%
  \BibitemOpen
  \bibfield{author}{%
  \bibinfo {author} {\bibfnamefont{L.}~\bibnamefont{B{\'a}nyai}}, \bibinfo
  {author} {\bibfnamefont{I.}~\bibnamefont{Galbraith}}, \bibinfo {author}
  {\bibfnamefont{C.}~\bibnamefont{Ell}},\ and\ \bibinfo {author}
  {\bibfnamefont{H.}~\bibnamefont{Haug}},\ }%
  \bibfield{journal}{%
  \bibinfo {journal} {Phys. Rev. B}\ }%
  \textbf{\bibinfo {volume} {36}},\ \bibinfo {pages} {6099} (\bibinfo {year}
  {1987})%
  \bibAnnoteFile{NoStop}{banyaigeh87}%
\bibitem{xie01}%
  \BibitemOpen
  \bibfield{author}{%
  \bibinfo {author} {\bibfnamefont{W.}~\bibnamefont{Xie}},\ }%
  \bibfield{journal}{%
  \bibinfo {journal} {J. Phys.: Condens. Matter}\ }%
  \textbf{\bibinfo {volume} {13}},\ \bibinfo {pages} {3149} (\bibinfo {year}
  {2001})%
  \bibAnnoteFile{NoStop}{xie01}%
\bibitem{nakaoka}%
  \BibitemOpen
  \bibfield{author}{%
  \bibinfo {author} {\bibfnamefont{T.}~\bibnamefont{Nakaoka}}, \bibinfo
  {author} {\bibfnamefont{T.}~\bibnamefont{Saito}}, \bibinfo {author}
  {\bibfnamefont{J.}~\bibnamefont{Tatebayashi}},\ and\ \bibinfo {author}
  {\bibfnamefont{Y.}~\bibnamefont{Arakawa}},\ }%
  \bibfield{journal}{%
  \bibinfo {journal} {Phys. Rev. B}\ }%
  \textbf{\bibinfo {volume} {70}},\ \bibinfo {pages} {235337} (\bibinfo {year}
  {2004})%
  \bibAnnoteFile{NoStop}{nakaoka}%
\bibitem{jovanov}%
  \BibitemOpen
  \bibfield{author}{%
  \bibinfo {author} {\bibfnamefont{V.}~\bibnamefont{Jovanov}}, \bibinfo
  {author} {\bibfnamefont{T.}~\bibnamefont{Eissfeller}}, \bibinfo {author}
  {\bibfnamefont{S.}~\bibnamefont{Kapfinger}}, \bibinfo {author}
  {\bibfnamefont{E.~C.}\ \bibnamefont{Clark}}, \bibinfo {author}
  {\bibfnamefont{F.}~\bibnamefont{Klotz}}, \bibinfo {author}
  {\bibfnamefont{M.}~\bibnamefont{Bichler}}, \bibinfo {author}
  {\bibfnamefont{J.~G.}\ \bibnamefont{Keizer}}, \bibinfo {author}
  {\bibfnamefont{P.~M.}\ \bibnamefont{Koenraad}}, \bibinfo {author}
  {\bibfnamefont{G.}~\bibnamefont{Abstreiter}},\ and\ \bibinfo {author}
  {\bibfnamefont{J.~J.}\ \bibnamefont{Finley}},\ }%
  \bibfield{journal}{%
  \bibinfo {journal} {Phys. Rev. B}\ }%
  \textbf{\bibinfo {volume} {83}},\ \bibinfo {pages} {161303} (\bibinfo {year} {2011})%
  \bibAnnoteFile{NoStop}{jovanov}%
\bibitem{wetting}%
  \BibitemOpen
  \bibfield{author}{%
  \bibinfo {author} {\bibfnamefont{T.}~\bibnamefont{Kazimierczuk}}, \bibinfo
  {author} {\bibfnamefont{A.}~\bibnamefont{Golnik}}, \bibinfo {author}
  {\bibfnamefont{P.}~\bibnamefont{Kossacki}}, \bibinfo {author}
  {\bibfnamefont{J.~A.}\ \bibnamefont{Gaj}}, \bibinfo {author}
  {\bibfnamefont{Z.~R.}\ \bibnamefont{Wasilewski}},\ and\ \bibinfo {author}
  {\bibfnamefont{A.}~\bibnamefont{Babi\ifmmode~\acute{n}\else \'{n}\fi{}ski}},\
  }%
  \bibfield{journal}{%
  \bibinfo {journal} {Phys. Rev. B}\ }%
  \textbf{\bibinfo {volume} {84}},\ \bibinfo {pages} {115325} (\bibinfo {year}
  {2011})%
  \bibAnnoteFile{NoStop}{wetting}%
\bibitem{nozik2008multiple}%
  \BibitemOpen
  \bibfield{author}{%
  \bibinfo {author} {\bibfnamefont{A.~J.}\ \bibnamefont{Nozik}},\ }%
  \bibfield{journal}{%
  \bibinfo {journal} {Chem. Phys. Lett.}\ }%
  \textbf{\bibinfo {volume} {457}},\ \bibinfo {pages} {3} (\bibinfo {year}
  {2008})%
  \bibAnnoteFile{NoStop}{nozik2008multiple}%
\bibitem{witzel}%
  \BibitemOpen
  \bibfield{author}{%
  \bibinfo {author} {\bibfnamefont{W.~M.}\ \bibnamefont{Witzel}}, \bibinfo
  {author} {\bibfnamefont{A.}~\bibnamefont{Shabaev}}, \bibinfo {author}
  {\bibfnamefont{C.~S.}\ \bibnamefont{Hellberg}}, \bibinfo {author}
  {\bibfnamefont{V.~L.}\ \bibnamefont{Jacobs}},\ and\ \bibinfo {author}
  {\bibfnamefont{A.~L.}\ \bibnamefont{Efros}},\ }%
  \bibfield{journal}{%
  \bibinfo {journal} {Phys. Rev. Lett.}\ }%
  \textbf{\bibinfo {volume} {105}},\ \bibinfo {pages} {137401} (\bibinfo {year}
  {2010})%
  \bibAnnoteFile{NoStop}{witzel}%
\bibitem{szimmermann08}%
  \BibitemOpen
  \bibfield{author}{%
  \bibinfo {author} {\bibfnamefont{C.}~\bibnamefont{Schindler}}\ and\ \bibinfo
  {author} {\bibfnamefont{R.}~\bibnamefont{Zimmermann}},\ }%
  \bibfield{journal}{%
  \bibinfo {journal} {Phys. Rev. B}\ }%
  \textbf{\bibinfo {volume} {78}},\ \bibinfo {pages} {045313} (\bibinfo {year}
  {2008})%
  \bibAnnoteFile{NoStop}{szimmermann08}%
\bibitem{mfogler08}%
  \BibitemOpen
  \bibfield{author}{%
  \bibinfo {author} {\bibfnamefont{A.~D.}\ \bibnamefont{Meyertholen}}\ and\
  \bibinfo {author} {\bibfnamefont{M.~M.}\ \bibnamefont{Fogler}},\ }%
  \bibfield{journal}{%
  \bibinfo {journal} {Phys. Rev. B}\ }%
  \textbf{\bibinfo {volume} {78}},\ \bibinfo {pages} {235307} (\bibinfo {year}
  {2008})%
  \bibAnnoteFile{NoStop}{mfogler08}%
\end{thebibliography}
\end{document}